\documentclass[twoside,twocolumn,9pt]{article}
\usepackage{verbatim}
\usepackage{extsizes}
\usepackage[super,sort&compress,comma]{natbib} 
\usepackage[version=3]{mhchem}
\usepackage[left=1.5cm, right=1.5cm, top=1.785cm, bottom=2.0cm]{geometry}
\usepackage{balance}
\usepackage{mathptmx}
\usepackage{sectsty}
\usepackage{graphicx} 
\usepackage{lastpage}
\usepackage{tikz}
\usetikzlibrary{intersections,arrows.meta} 
\usepackage[format=plain,justification=justified,singlelinecheck=false,font={stretch=1.125,small,sf},labelfont=bf,labelsep=space]{caption}
\usepackage{float}
\usepackage{fancyhdr}
\usepackage{fnpos}
\usepackage[english]{babel}
\addto{\captionsenglish}{%
  
}
\usepackage{array}
\usepackage{droidsans}
\usepackage{charter}
\usepackage[T1]{fontenc}
\usepackage[usenames,dvipsnames]{xcolor}
\usepackage{setspace}
\usepackage[compact]{titlesec}
\usepackage{hyperref}

\usepackage{epstopdf}

\graphicspath{{figs/}}

\newcommand{\etalia}{{\it et al.}}
\newcommand{\la}{\left\langle}
\newcommand{\ra}{\right\rangle}

\definecolor{cream}{RGB}{222,217,201}

\begin{document}

\pagestyle{fancy}
\thispagestyle{plain}
\fancypagestyle{plain}{
\renewcommand{\headrulewidth}{0pt}
}

\makeFNbottom
\makeatletter
\renewcommand\LARGE{\@setfontsize\LARGE{15pt}{17}}
\renewcommand\Large{\@setfontsize\Large{12pt}{14}}
\renewcommand\large{\@setfontsize\large{10pt}{12}}
\renewcommand\footnotesize{\@setfontsize\footnotesize{7pt}{10}}
\makeatother

\renewcommand{\thefootnote}{\fnsymbol{footnote}}
\renewcommand\footnoterule{\vspace*{1pt}%
\color{cream}\hrule width 3.5in height 0.4pt \color{black}\vspace*{5pt}} 
\setcounter{secnumdepth}{5}

\makeatletter 
\renewcommand\@biblabel[1]{#1}            
\renewcommand\@makefntext[1]%
{\noindent\makebox[0pt][r]{\@thefnmark\,}#1}
\makeatother 
\renewcommand{\figurename}{\small{Fig.}~}
\sectionfont{\sffamily\Large}
\subsectionfont{\normalsize}
\subsubsectionfont{\bf}
\setstretch{1.125} 
\setlength{\skip\footins}{0.8cm}
\setlength{\footnotesep}{0.25cm}
\setlength{\jot}{10pt}
\titlespacing*{\section}{0pt}{4pt}{4pt}
\titlespacing*{\subsection}{0pt}{15pt}{1pt}

\fancyfoot{}
\fancyfoot[LO,RE]{\vspace{-7.1pt}\includegraphics[height=9pt]{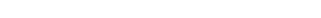}}
\fancyfoot[CO]{\vspace{-7.1pt}\hspace{13.2cm}\includegraphics{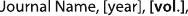}}
\fancyfoot[CE]{\vspace{-7.2pt}\hspace{-14.2cm}\includegraphics{head_foot/RF}}
\fancyfoot[RO]{\footnotesize{\sffamily{1--\pageref{LastPage} ~\textbar  \hspace{2pt}\thepage}}}
\fancyfoot[LE]{\footnotesize{\sffamily{\thepage~\textbar\hspace{3.45cm} 1--\pageref{LastPage}}}}
\fancyhead{}
\renewcommand{\headrulewidth}{0pt} 
\renewcommand{\footrulewidth}{0pt}
\setlength{\arrayrulewidth}{1pt}
\setlength{\columnsep}{6.5mm}
\setlength\bibsep{1pt}

\makeatletter 
\newlength{\figrulesep} 
\setlength{\figrulesep}{0.5\textfloatsep} 

\newcommand{\topfigrule}{\vspace*{-1pt}%
\noindent{\color{cream}\rule[-\figrulesep]{\columnwidth}{1.5pt}} }

\newcommand{\botfigrule}{\vspace*{-2pt}%
\noindent{\color{cream}\rule[\figrulesep]{\columnwidth}{1.5pt}} }

\newcommand{\dblfigrule}{\vspace*{-1pt}%
\noindent{\color{cream}\rule[-\figrulesep]{\textwidth}{1.5pt}} }

\makeatother

\twocolumn[
  \begin{@twocolumnfalse}
{\includegraphics[height=30pt]{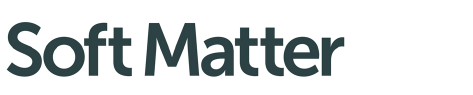}\hfill\raisebox{0pt}[0pt][0pt]{\includegraphics[height=55pt]{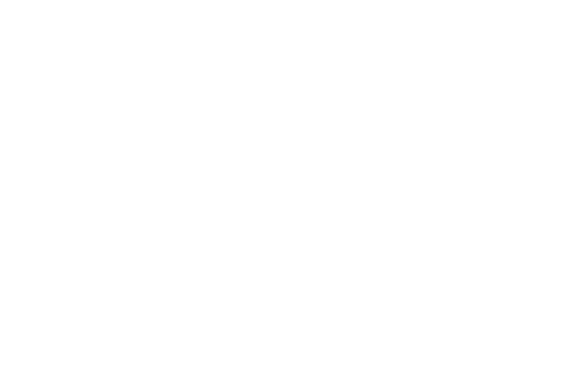}}\\[1ex]
\includegraphics[width=18.5cm]{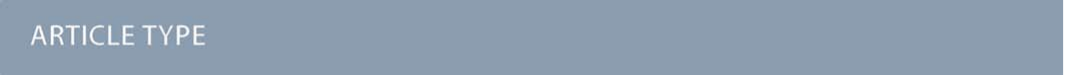}}\par
\vspace{1em}
\sffamily
\begin{tabular}{m{4.5cm} p{13.5cm} }

\includegraphics{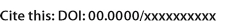} & 
\noindent\LARGE{\textbf{Modeling Crystallization in Concentrated Suspensions of Compressible Microgels}} 
\\
\vspace{0.3cm} & \vspace{0.3cm} \\

& \noindent\large{Oreoluwa E. Alade and Alan R. Denton\textit{$^{\ast}$}} \\

\includegraphics{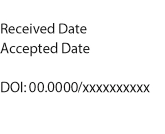} & \noindent\normalsize{Soft colloidal particles, such as microgels, made of cross-linked polymer gels that can swell by absorbing a good solvent, have internal degrees of freedom that allow them to respond to external stimuli by changing size. Sensitive responses to variations in temperature and concentration have inspired many practical applications, e.g., to drug delivery, biosensing, filtration, and photonic crystals. At sufficiently high concentrations, mutual crowding can drive microgels to interpenetrate or deform by faceting. To explore the influence of particle softness and compressibility on thermodynamic phase stability, we develop a coarse-grained model in which spherical microgels interact via the Hertz elastic pair potential and can respond to crowding by deswelling and faceting or interpenetrating, as governed by the Flory-Rehner theory of polymer networks. In our model, faceting reduces the volume available for swelling, while interpenetration affects polymer-solvent mixing entropy. By performing Monte Carlo simulations that incorporate novel trial changes in particle size and shape, we compute 
equilibrium swelling ratios and
Lindemann ratios to determine the fluid-solid phase stability boundary over a range of particle softness, tuned by varying cross-link fraction. 
Predictions from our facet model are in qualitative agreement with experiments and molecular-scale simulations.
In contrast to the phase behavior of incompressible Hertzian spheres, we find that suspensions of compressible microgels crystallize at significantly lower volume fractions when accounting for the free energy costs of faceting and interpenetration. Our results extend previous studies of microgel phase behavior and can help to interpret experiments.} \\

\end{tabular}

 \end{@twocolumnfalse} \vspace{0.6cm}

  ]
  
\renewcommand*\rmdefault{bch}\normalfont\upshape
\rmfamily
\section*{}
\vspace{-1cm}


\footnotetext{\textit{Department of Physics, North Dakota State University, Fargo, ND 58108-6050, USA. $^*$E-mail: alan.denton@ndsu.edu}}

\section{Introduction}\label{introduction}

Microgels are soft colloidal hydrogel particles composed of cross-linked polymer networks that can swell in solution by absorbing solvent.\cite{Pelton2000,MicrogelBook2011,HydrogelBook2012} Their internal degrees of freedom allow them to adjust size in response to changes in external stimuli, such as temperature or concentration. This responsiveness underpins diverse applications, including drug delivery, biosensing, filtration, and photonic crystals.\cite{MicrogelBook2011,LyonFernandezNieves2012,Yunker2014} At high concentrations, microgels exhibit rich thermodynamic behavior, including crystallization, solid-solid phase transitions, and even reentrant melting.\cite{Senff1999,Stieger2004}

Extensive experimental and theoretical studies have examined the elasticity of individual microgels~\cite{Borrega1999,Cloitre2003,Tan2004,FernandezNieves2000,FernandezNieves2003,LietorSantos2009,Hertle2010,Menut2012,Romeo2012,Romeo2013,LietorSantos2011SM,SierraMartin2012,LietorSantos2011PRE,SierraMartin2011,Hashmi2009,PelaezFernandez2015} and the equilibrium and dynamical properties of bulk suspensions.\cite{Mason1995,Grohn2000,Levin2002,FernandezNieves2005,Singh2012,Winkler2014,Li2014,Egorov2013,Colla2014,Colla2015,Gupta2015Nanoscale,Gupta2015PRL,scotti2022} Links between single-particle properties, such as cross-link fraction, and collective properties, including osmotic pressure, phase transitions, pair correlations, and viscosity, have been established using techniques such as light scattering, neutron scattering, confocal microscopy, and osmometry. While swelling behavior is now well documented, the broader consequences of elasticity and compressibility for bulk properties remain unresolved.

Combining osmometry and dialysis, Pelaez-Fernandez \textit{et al.}\cite{PelaezFernandez2015} showed that 
osmotic pressure in suspensions of ionic poly(4-vinylpyridine) (PVP) microgels 
is dominated by counterions, allowing accurate determination of volume fractions up to high densities. Their results demonstrated that increasing particle softness narrows the fluid-solid coexistence gap or suppresses crystallization entirely, in agreement with Hertzian and penetrable-sphere models.\cite{Pamies2009Hertzian,Likos1998Penetrable} More recently, Hildebrandt \textit{et al.}\cite{Hildebrandt2023FluidSolid} investigated thermoresponsive poly(N-isopropylacrylamide) (PNIPAM) core-shell microgels and showed that temperature controls crystallization into long-range-ordered photonic crystals as well as remelting into the fluid phase. Their measurements further revealed continuous particle shrinking with increasing temperature and strong deviations from hard-sphere behavior at high concentrations, underscoring the role of deformability in phase stability. 

Imaging studies reveal that microgels in concentrated suspensions may respond to mutual crowding 
by interpenetrating, deforming, and compressing. From superresolution microscopy measurements, 
Conley \textit{et al.}\cite{Conley2017Jamming,conley2019,scheffold2020pathways} identified 
three stages of response with increasing concentration. In the first stage, at relatively 
low packing fraction, no deformation is visible as the particles are mildly compressed and begin 
to interpenetrate. However, considering that microgels typically have diffuse peripheries, consisting
of a fuzzy corona and dangling polymer chains surrounding a dense core, their dSTORM method is not 
sensitive enough to detect possible weak deformation of coronas. In the second stage, with increasing 
packing fraction, interpenetration continues while the particles start to facet as their cores 
make contact. In the third stage, at the highest packing fractions, faceting and interpenetration 
finally saturate and the cores start to deswell.

Complementary molecular-scale modeling by Zaccarelli {\it et al.},\cite{zaccarelli-multiHertzian2018,
Rovigatti2019Hertzian,Rovigatti2019overview,zaccarelli2024} who performed molecular dynamics simulations 
of microgels modeled as bead-spring polymer networks, have tested the validity of the Hertzian model 
of elastic interparticle interactions
and connected single-particle properties to collective behavior of microgel suspensions.
From extensive simulations of suspensions of many (108) interacting microgels, each comprising 5000 beads, 
in an implicit solvent, Del Monte and Zaccarelli~\cite{zaccarelli2024} showed that microgels 
with an internal core-shell structure similar to that synthesized in experiments responded to 
increasing concentration in a manner qualitatively similar to that observed by 
Conley \textit{et al.}\cite{Conley2017Jamming,conley2019}

A key step in modeling thermodynamic phase behavior of bulk suspensions of soft colloids was taken by P{\`a}mies \textit{et al.},\cite{Pamies2009Hertzian} who studied incompressible Hertzian spheres to assess the effect of elastic interactions on phase transitions. Using Monte Carlo simulations and thermodynamic integration,\cite{FrenkelSmit2023,FrenkelLadd1984} they computed free energies of fluid and solid phases and analyzed phase coexistence. Their results revealed a fluid-FCC transition, followed by FCC-BCC and further solid-solid transformations at higher densities and low temperatures. Subsequent work by Munaò and Saija,\cite{MunaoSaija2019Hertzian} employing Monte Carlo simulations and Ornstein-Zernike integral equation theory, confirmed these findings and highlighted the reentrant melting behavior characteristic of the Hertzian model.
The phase diagram and melting transition of two-dimensional Hertzian discs has also been studied.
\cite{Fomin2018Hertzian2D,Tsiok2020Hertzian2D} 

Building on these foundations, we extend the Hertzian framework to account for microgel compressibility and deformability. In our approach, particles can swell or deswell, interpenetrate or deform by faceting in response to crowding, thus incorporating crucial internal degrees of freedom absent from incompressible particle models. Using Monte Carlo simulations, we compute Lindemann ratios, swelling ratios, and structural properties of fluid and solid phases to assess phase stability. Our goal is to provide a more realistic description of microgel suspensions at high concentrations, where softness, interpenetration, and deformation govern fluid-solid transitions.

The remainder of the paper is organized as follows. 
In Sec.~\ref{models}, we describe coarse-grained models of compressible microgels that couple the Flory-Rehner theory of polymer networks with the Hertz theory of elastic pair interactions and introduce extensions that account for faceting and interpenetration. Section~\ref{methods} describes our Monte Carlo simulation methods and the Lindemann criterion used to identify fluid-solid transitions. In Sec.~\ref{results}, we report simulation results for the equilibrium Lindemann ratio, swelling behavior, and phase stability, and we construct fluid-solid stability curves in the plane of volume fraction and particle softness. In the process, we validate the facet model against theoretical calculations. Finally, Sec.~\ref{conclusions} summarizes our findings and discusses implications for experiments on microgel suspensions.

\section{Models and Theories}\label{models}

\subsection{Coarse-Grained Model of Microgels}

The system of interest is a suspension of $N_m$ microgels dispersed in a solvent at constant temperature $T$. Each microgel comprises $N_{\rm mon}$ polymer monomers bonded together to form $N_{\rm ch}$ linear homopolymer chains cross-linked into a network. For tetravalent cross-linkers, the cross-link fraction is $x=0.5N_{\rm ch}/N_{\rm mon}$. In its collapsed state, each microgel has a dry radius $a_0$, and when swollen by solvent a radius $a$. If monomers of radius $r_{\rm mon}$ are randomly close-packed in the dry state, then $N_{\rm mon}\simeq 0.63(a_0/r_{\rm mon})^3$. The degree of swelling can be characterized by the linear swelling ratio $\alpha\equiv a/a_0$ or, alternatively, the polymer (monomer) volume fraction $\phi_p=1/\alpha^3$. Mass conservation dictates 
a solvent volume fraction $\phi_s=1-\phi_p$. Typically, $\phi_p\ll 1$, implying that swollen microgels contain mostly solvent.
As is common in modeling polymeric gels, we assume swelling of the network to be affine (uniform and isotropic). 
Although microgels often comprise a densely cross-linked core surrounded by a softer shell,\cite{zaccarelli2024,moncho-jorda2025} we consider here homogeneous microgels with a uniform distribution of cross-links, which have been experimentally synthesized.\cite{pelton2011,varga2011,karg-richtering2019}
For typical system parameters, the total number of particles (polymer monomers plus solvent molecules) within even a single microgel can easily exceed $N_{\rm mon}\alpha^3\sim{\cal O}(10^8)$.
Therefore, for computational efficiency, we consider a practical coarse-grained model, which greatly simplifies the representation of such a complex system by retaining only essential degrees of freedom, while averaging out those deemed less relevant. This approach reduces the complexity of a swollen microgel while preserving the key physical features of interest. Following previous work of Denton \etalia,\cite{UrichDenton2016,WeyerDenton2018Swelling} we combine the Flory-Rehner theory of polymer networks with Hertzian elastic pair interactions to model swelling properties and phase stability of nonionic microgels, ignoring electrostatic interactions.\cite{WeyerDenton2018Swelling}
However, we also consider the influence of faceting (Sec.~\ref{facet_model_section}) and 
interpenetration (Sec.~\ref{penetration_model_section}) of microgels. 

Instead of explicitly modeling all polymer monomers and solvent molecules, the coarse-grained approach averages over most degrees of freedom of these microscopic constituents, allowing microgels to be treated as continuum particles, immersed in an implicit solvent, and governed by an effective Hamiltonian. In the canonical ensemble, the partition function is given by
\begin{equation}\label{partition_function}
\mathcal{Z} = \Big\langle \Big\langle e^{-\beta (H_p + H_m)} \Big\rangle_m \Big\rangle_p,
\end{equation}
where $\beta\equiv 1/(k_BT)$ and angular brackets denote 
statistical traces (unnormalized ensemble averages) 
over internal polymeric ($p$) and external microgel ($m$) center-of-mass degrees of freedom. 
For example,
\begin{equation}\label{tracem}
\Big\langle~\Big\rangle_m = \frac{1}{{N_m}!\Lambda^{3N_m}}\int d{\bf r}_1\cdots\int d{\bf r}_{N_m},
\end{equation}
where $\{{\bf r}_1,\ldots,{\bf r}_{N_m}\}$ are the microgel positions and $\Lambda$ is the microgel 
thermal de Broglie wavelength. In Eq.~(\ref{partition_function}), $H_p$ is the polymeric Hamiltonian, 
associated with microscopic degrees of freedom of the polymer chains, while $H_m$ is the microgel Hamiltonian, 
which combines the total kinetic and potential energies.

If we assume only pairwise interparticle interactions, the Hamiltonian can be separated into 
one-body (self) and pair terms:
\begin{equation}
H=H_{\rm self}+H_{\rm pair}.
\label{H}
\end{equation}
It should be noted that the pairwise assumption neglects many-body interactions, which may be significant 
in dense suspensions. If we further assume that swelling is governed only by a subset of polymeric degrees 
of freedom (collectively denoted by $a$), and denote the remaining internal degrees of freedom by $p'$, 
the partition function may be written as
\begin{equation}
{\cal Z}=\la\la e^{-\beta H_{\rm eff}}\ra_m\ra_a,
\label{Z2}
\end{equation}
where $\la~\ra_a$ denotes a trace over all microgel radii and
\begin{equation}
H_{\rm eff}=-k_BT\ln\la e^{-\beta(H_{\rm self}+H_{\rm pair})}\ra_{p'}
\label{Heff}
\end{equation}
is an effective Hamiltonian resulting from tracing over the $p'$ internal polymeric coordinates.
Now making a mean-field approximation that neglects correlations between microgel swelling and
pair interactions, and thus decouples one-body and pair coordinates, we write
\begin{equation}
H_{\rm eff}=-k_BT\ln\left(\la e^{-\beta H_{\rm self}}\ra_{p'}\la e^{-\beta H_{\rm pair}}\ra_{p'}\right)
=F_{\rm self}+U_{\rm eff},
\label{Heff2}
\end{equation}
thus defining a single-microgel (self) polymeric free energy,
\begin{equation}
F_{\rm self}=-k_BT\ln\la e^{-\beta H_{\rm self}}\ra_{p'}=\sum_{i=1}^{N_m}F_{{\rm self},i}(a_i),
\label{Fself}
\end{equation}
which depends only on the microgel swollen radii, and an effective pair interaction energy,
\begin{equation}
U_{\rm eff}=-k_BT\ln\la e^{-\beta H_{\rm pair}}\ra_{p'}=\sum_{i<j=1}^{N_m}v_{\rm eff}(r_{ij}),
\label{Ueff}
\end{equation}
where $v_{\rm eff}(r_{ij})$ is an effective pair potential, dependent on the microgel swollen radii,
describing the interaction energy between two microgels, separated by distance $r_{ij}$, and whose explicit form
depends on the model, as described below in Secs.~\ref{Hertz_model_section}-\ref{penetration_model_section}.

Thus, in this coarse-grained model, after tracing over a subset of internal degrees of freedom of 
the microgels, the canonical partition function reduces to
\begin{equation}
{\cal Z}=\la e^{-\beta F_{\rm self}(a)}\la e^{-\beta U_\text{eff}(a)}\ra_m\ra_a.
\label{Z3}
\end{equation}
Next, we approximate $F_{\rm self}$ by appealing to Flory-Rehner theory and we relate $U_{\rm eff}$ 
to the Hertzian model and two alternative models, one of which also incorporates many-body interactions.

\subsection{Flory-Rehner Theory of Microgels}\label{FR theory}

The equilibrium swelling properties of polymer networks can be described by the classical Flory-Rehner theory,\cite{FloryRehner1943I,FloryRehner1943II,Flory1953Principles} which combines mixing entropy, conformational
entropy, and polymer-solvent interaction energy. Although originally formulated for macroscopic gels, 
this framework has been widely adapted to microgels and qualitatively captures their swelling behavior, 
albeit with certain quantitative limitations.\cite{FernandezNieves2003,Riest2015SciRep} In $k_BT$ units, 
the free energy of a microgel, which may be identified with $F_{\rm self}$ in Eq.~(\ref{Fself}),
is given by 
\begin{eqnarray}
\beta F_{\text{gel}}(\alpha) &=& N_{\rm mon}\left[ (\alpha^3 - 1)\ln(1-\alpha^{-3}) 
+ \chi (1-\alpha^{-3}) \right] 
\nonumber\\[1ex]
&+& \tfrac{3}{2} N_{\rm ch} (\alpha^2 - 1 - \ln \alpha).
\label{eq:flory-rehner-equation}
\end{eqnarray}
Here \(\chi\) denotes the Flory solvency parameter, which depends, in a mean-field approximation, on mean polymer-polymer, polymer-solvent, and solvent-solvent interaction energies.\cite{doi2013soft} The first bracketed term represents the entropy of mixing together with polymer-solvent interaction energy, while the last term accounts for the conformational entropy of the isotropic network. The theory assumes Gaussian chain statistics,\cite{doi2013soft,rubinstein-colby2023} which is appropriate when chain end-to-end distances remain much smaller than the contour length,\cite{Rumyantsev2015} i.e., for swelling ratios satisfying \(\alpha \ll 1/\sqrt{x}\). The polymeric free energy in Eq.~(\ref{Fself}) is the sum over microgels of Eq.~(\ref{eq:flory-rehner-equation}).

Extensive studies~\cite{Pelton2000,Saunders2009,FernandezNieves2000,FernandezNieves2003,LietorSantos2009,Hertle2010,Menut2012,Romeo2012,Romeo2013,LietorSantos2011SM,Riest2012,SierraMartin2012,LietorSantos2011PRE,SierraMartin2011} have demonstrated that the Flory-Rehner model provides a useful description of the swelling of loosely cross-linked microgels, even though it systematically overestimates the solvency parameter.\cite{LopezRichtering2017} More advanced approaches, which incorporate network heterogeneity and non-Gaussian conformations, can offer improved quantitative accuracy.\cite{BlundellTerentjev2009,Roa2017Catalysis}

\subsection{Hertzian Model of Microgel Interactions}\label{Hertz_model_section}

\begin{figure}[h]
\centering
\includegraphics[width=0.9\linewidth]{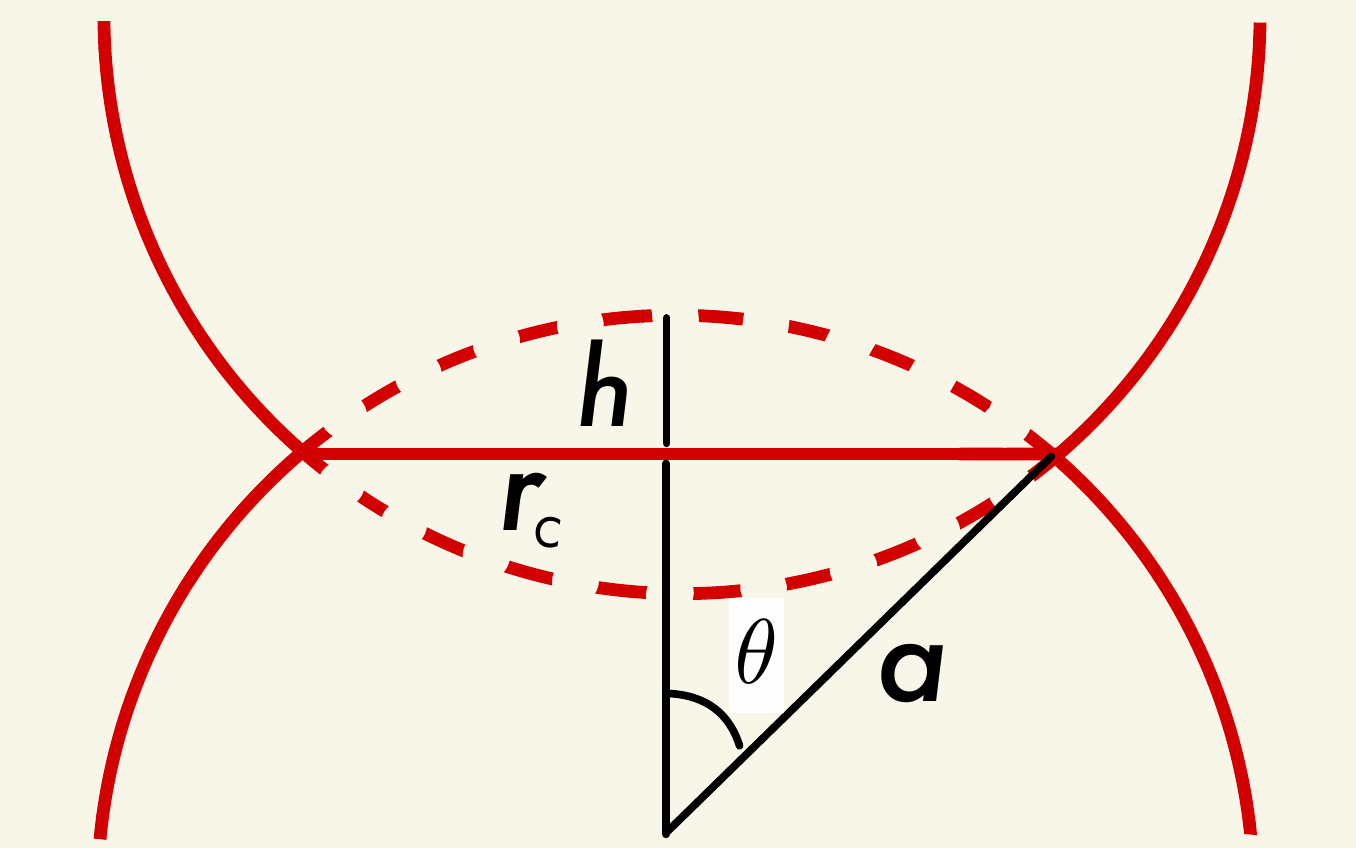}
\caption{
Schematic drawing of two elastic spheres of radius $a$ that, upon contact, flatten to form 
spherical caps of height $h$ and a disc-shaped facet of radius $r_c$. In the Hertz contact theory,
$h\ll r_c\ll a$.}
\label{fig:hertz_model}
\end{figure}

Within a conceptual framework that idealizes microgels as solid elastic spheres, interparticle 
interactions have been widely described by the Hertz contact theory and variations thereof.
\cite{zaccarelli-multiHertzian2018,moncho-jorda2025} When two solid spheres come into contact, they may 
deform elastically by forming flat spherical caps, resulting in a short-ranged repulsive force that 
strengthens with increasing compression. The original Hertz contact theory~\cite{Hertz1881contact,
Hertz1896contact,Hertz1882hardness,Hertz1896hardness} predicts the associated elastic energy resulting from 
the strain, which is assumed to be localized near contact.\cite{LandauLifshitzElasticity1986} 
As depicted in Fig.~\ref{fig:hertz_model}, for spheres of equal radius $a$ compressed to a depth 
(cap height) $h\ll a$, the size of the perturbed (strained) volume is assumed to be comparable to the 
radius $r_c$ of the contact disc.\cite{AudolyPomeau2010} Assuming weak deformation, such that 
$h\ll r_c\ll a$, the strain (gradient of displacement) is uniform and of order $h/r_c$. Furthermore, 
if $\theta$ is the angle subtended at the center of the sphere by the disc radius, then
\begin{equation}\label{eq:hertz_model}
\cos\theta \simeq 1-\frac{1}{2}\left(\frac{r_c}{a}\right)^2 \simeq \frac{a-h}{a},
\end{equation}
which implies $r_c\sim(ha)^{1/2}$. Within a Hookean elasticity assumption, valid for small strains,
the elastic energy is proportional to the product of Young's modulus $Y$, 
the strained volume ($\sim r_c^3$), and the square of the strain $(h/r_c)^2$, thus scaling as $\sim h^{5/2}$. 

For a pair of particles, $i$ and $j$, of radii $a_i$ and $a_j$, with center-to-center separation 
$r_{ij}=a_i+a_j-h$, the Hertz pair potential is~\cite{LandauLifshitzElasticity1986} 
\begin{equation}\label{eq:Hertz_pair_potential}
  v_H(r_{ij}) =
  \begin{cases}
    \varepsilon \left( 1 - \dfrac{r_{ij}}{a_i + a_j} \right)^{5/2}, & r_{ij} < a_i + a_j, \\[2ex]
    0, & r_{ij} \geq a_i + a_j,
  \end{cases}
\end{equation}
with amplitude $\varepsilon$. Applying the Hertzian model to a suspension of $N_m$ microgels, 
assuming pairwise interactions, the total elastic interaction energy is
\begin{equation}\label{eq:internal_energy}
    U_H = \sum_{i<j=1}^{N_m} v_H(r_{ij}),
\end{equation}
which may be identified with the effective pair interaction energy in Eq.~(\ref{Ueff}).
The amplitude $\varepsilon$ in Eq.~(\ref{eq:Hertz_pair_potential}) describes the resistance of the particles 
to elastic compression. For uniform spheres (uniformly cross-linked microgels) of Young’s moduli $Y_i$ and 
$Y_j$ and Poisson’s ratios $\nu_i$ and $\nu_j$, the Hertz amplitude is 
given by~\cite{LandauLifshitzElasticity1986,WeyerDenton2018Swelling}
\begin{equation}\label{epsilon_general}
    \varepsilon = \frac{8}{15} \left( \frac{1 - \nu_i^2}{Y_i} + \frac{1 - \nu_j^2}{Y_j} \right)^{-1} (a_i + a_j)^2 \sqrt{a_i a_j}.
\end{equation}
If the particles are identical (\(a_i = a_j = a\), \(Y_i = Y_j = Y\), \(\nu_i = \nu_j = \nu\)), then Eq.~\eqref{epsilon_general} reduces to
\begin{equation}\label{epsilon}
    \varepsilon = \frac{4 Y V_m}{5\pi (1 - \nu^2)},
\end{equation}
where $V_m=(4/3)\pi a^3$ is the swollen microgel volume.\cite{UrichDenton2016}
For polymer gels in good solvents, typical Poisson's ratios are $\nu\simeq 0.5$.\cite{geissler1981,li1993,matzelle2003}
In the limit $Y\to\infty$, Eq.~\eqref{eq:Hertz_pair_potential} reduces to a hard-sphere pair potential,
\begin{equation}\label{eq:vhs}
  v_{\rm HS}(r) =
  \begin{cases}
    \infty, & r < a_i + a_j, \\[2ex]
    0, & r \geq a_i + a_j.
  \end{cases}
\end{equation}

For polymer gels in good solvents, scaling theory~\cite{deGennes1979Scaling} predicts
\begin{equation}\label{eq:Y_scaling}
    Y = \frac{Y_c k_B T N_{\text{ch}}}{a^3},
\end{equation}
where $Y_c$ is a dimensionless factor for calibrating models to experiments. This relation implies that microgels become stiffer with increasing cross-link fraction (higher $N_{\text{ch}}$) and smaller swollen size. Experimental studies support this picture: Li\'etor-Santos \textit{et al.}\cite{LietorSantos2011SM} reported Young’s moduli of $0.4-1.6$ kPa for PVP microgels, while Sierra-Martín \textit{et al.}\cite{SierraMartin2011} found PNIPAM microgels to exhibit sub-kilopascal bulk moduli and pronounced softening near the volume phase transition.

It is important to note that the Hertz potential for microgels is intrinsically coupled to the Flory-Rehner free energy through dependence of the amplitude [Eq.~(\ref{epsilon})] on cross-link fraction $x$ ($Y \propto x$), which governs both swelling and elasticity of microgels. In the Hertz theory, a higher cross-link fraction strengthens elastic repulsion between particles, while in the Flory-Rehner framework it suppresses swelling by limiting solvent uptake. The cross-link fraction thus provides a unifying parameter that controls both the mechanical stiffness and swelling response of microgels. In contrast, the classic Hertz theory for solid elastic spheres treats the Young's modulus as a fixed material property.

A key advantage of the Hertz theory is its computational efficiency, making it widely used in large-scale simulations of soft colloids.\cite{Pamies2009Hertzian} Recent extensions to a multi-Hertzian model~\cite{zaccarelli-multiHertzian2018,moncho-jorda2025} allow modeling microgels with a heterogeneous (e.g., core-shell) cross-link distribution. At high concentrations, however, microgels deform significantly and the pairwise Hertz theory breaks down. Simulations of a molecular model by Rovigatti \textit{et al.}\cite{Rovigatti2019Hertzian} showed that, while the Hertz potential is accurate at weak to moderate overlap, it fails to capture many-body deformations and anisotropies that dominate strongly overlapping microgels. This limitation motivates extensions that incorporate internal degrees of freedom, such as deswelling, faceting, and interpenetration.

Microgels are viscoelastic objects that exhibit properties characteristic of both viscous fluids and 
elastic solids. In response to mechanical stress, the solvent may flow in and out of the gel, affecting 
polymer-solvent mixing entropy, while the polymer network may elastically deform and swell or deswell, 
affecting conformational entropy, as described by the first and last terms, respectively, of the Flory-Rehner 
free energy [Eq.~(\ref{eq:flory-rehner-equation})]. Removing cross-linkers reduces the gel to a droplet
of dilute polymer solution, while adding cross-linkers and expelling solvent transforms the gel into an 
elastomeric, rubber-like ball. 

The Hertz theory reasonably captures the elastic response of microgels to localized contact and deformation, 
but neglects attendant changes in entropy due to inflow and drainage of solvent. In the next subsection, 
we introduce a complementary model that aims to incorporate, albeit in an idealized fashion, thermodynamic
effects of polymer and solvent redistribution that may occur in highly concentrated suspensions of microgels.

\subsection{Facet Model of Microgels}\label{facet_model_section}

\begin{figure}[h]
    \centering
    \includegraphics[width=0.9\linewidth]{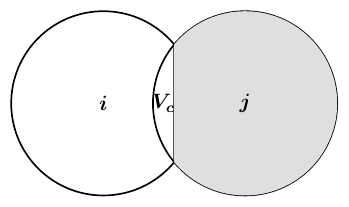}
\caption{Schematic drawing of two interacting microgels \(i\) and \(j\) that form a flat facet upon contact. The shaded region of microgel \(j\) depicts the volume remaining after removal of a spherical cap of volume \(V_c\), which increases the local polymer concentration, lowering the swelling ratio.}
\label{fig:facet_model}
\end{figure}

Given that microgels are not solid elastic spheres, whose deformations upon contact are governed only by 
elastic forces, but rather are polymeric colloidal particles dispersed in a solvent and influenced by 
polymer-solvent mixing entropy, the Hertz theory may not fully capture their response to crowding. 
Specifically, the theory does not account for changes in entropy resulting from internal redistribution 
of polymer and solvent.  Motivated by these considerations, as well as by 
the experiments of Conley \textit{et al.}\cite{Conley2017Jamming,conley2019,scheffold2020pathways}
and the simulations of Del Monte and Zaccarelli,\cite{zaccarelli2024} we consider
an extension of the Hertz model in which the facets formed by contacting microgels reduce the effective 
microgel volume by excluding polymer and solvent from the cap regions (see Fig.~\ref{fig:facet_model}).
For particles with internal (polymeric) degrees of freedom, this shape deformation redistributes polymer 
and solvent in the unfaceted volume (nonlocal response), thereby decreasing the swelling ratio and 
modifying the free energy. In Sec.~\ref{penetration_model_section} below, we consider a different 
physical model in which polymer networks of neighboring microgels may interpenetrate and share an 
overlap volume. 

Upon the faceting of two microgels, $i$ and $j$, of radii $a_i$ and $a_j$, whose centers are separated by a distance $r$, the volume of microgel $i$ decreases by the volume of a spherical cap,
\begin{equation}
V_{c,ij} = \frac{\pi}{3}h_{ij}^2(3a_i-h_{ij}),
\label{Vcij}
\end{equation}
with height
\begin{equation}
h_{ij} = \frac{(r-a_i+a_j)(a_i+a_j-r)}{2r}.
\label{hij}
\end{equation}
Summing over all neighboring microgels $j$ that form a facet with microgel $i$, 
and assuming a concentration low enough that neighboring facets on the same particle do not intersect,
the total volume of all caps on microgel $i$ is
\begin{equation}
V_{c,i} = \sum_j V_{c,ij}.
\label{Vctotal}
\end{equation}
Assuming that the polymer is uniformly redistributed in the reduced volume, $V_m-V_{c,i}$, the modified swelling ratio becomes
\begin{equation}\label{tilde_alpha1}
    \tilde{\alpha} = \alpha\left(1-\frac{V_{c,i}}{V_{m}} \right)^{1/3}.
\end{equation}
Substituting $\tilde{\alpha}$ into the Flory-Rehner expression yields a modified gel free energy,
\begin{eqnarray}
\beta\tilde F_\text{gel} &=& N_{\rm mon} \Big[ (\tilde{\alpha}^3 - 1)\ln(1 - \tilde{\alpha}^{-3}) 
+ \chi (1 - \tilde{\alpha}^{-3}) \Big] \nonumber\\[1ex]
&+& \tfrac{3}{2} N_{\rm ch} \big( \tilde{\alpha}^2 - 1 - \ln \tilde{\alpha} \big).
\label{modified_fr_energy}
\end{eqnarray}
In terms of the polymer volume fraction,
\begin{equation}\label{newphi_p}
    \tilde{\phi}_p = \phi_p \left(1 - \frac{V_{c,i}}{V_m}\right)^{-1} = \tilde{\alpha}^{-3},
\end{equation}
the modified free energy may be equivalently written as
\begin{eqnarray}\label{nonlocalfacet}  
\beta\tilde F_\text{gel} &=& N_{\rm mon} \Big[ (\tilde{\phi}_p^{-1} - 1)\ln(1 - \tilde{\phi}_p) 
+ \chi (1 - \tilde{\phi}_p) \Big] \nonumber\\[1ex]
&+& \tfrac{3}{2}N_{\rm ch} \Big( \tilde{\phi}_p^{-2/3} - 1 + \tfrac{1}{3}\ln\tilde{\phi}_p \Big).
\end{eqnarray}
The total free energy of the system, combining gel swelling free energy and elastic pair interaction energy, 
is then given by
\begin{equation}\label{F_fac}
    F_\text{fac} = \sum_{i=1}^{N_m} \tilde F_\text{gel}(\tilde{\alpha}_i) + \sum_{i<j=1}^{N_m}v_H(r_{ij}; \alpha_i, \alpha_j).
\end{equation}

Supplementing
the Hertzian model, our facet model accounts for the redistribution of polymer and solvent 
away from the contact region and into a reduced (unfaceted) volume, which modifies the polymer-solvent 
mixing entropy and the polymer conformational entropy in the swelling free energy 
[Eq.~(\ref{modified_fr_energy}) or (\ref{nonlocalfacet})]. 
The influence of facets thus extends beyond the local region of contact, propagating throughout the 
microgel volume. 
For the sake of simplicity, we assume here that in the reduced volume the polymer and cross-linkers 
remain uniformly distributed and swelling of the network remains affine. 

Whether the facet model should supplement or rather replace the Hertz model is a legitimate question,
considering that the Flory-Rehner theory includes polymer conformational entropy, which might be 
reasonably presumed the origin of a microgel's elastic response to contact. Here, we choose to combine 
the two models on the grounds that deformation of contacting microgels is localized near the contact region,
and thus best described by the Hookean spring energy intrinsic to the Hertz potential, with an amplitude 
determined by the Young's modulus via the cross-link density, while the resulting redistribution of 
polymer and solvent of the entire microgel into the unfaceted volume is a nonlocal response best 
described by the Flory-Rehner entropic free energy with modified swelling ratio.

Note that $\tilde F_\text{gel}$ in Eqs.~\eqref{modified_fr_energy} and \eqref{nonlocalfacet} may depend on the deformed shapes and resulting polymer redistribution of multiple faceted microgels (through $\tilde\alpha$), and thus is inherently a many-body free energy, especially at high concentrations. In contrast, the original Flory-Rehner free energy [Eq.~\eqref{eq:flory-rehner-equation}] is a one-body property, depending on the swelling ratio of only a single microgel, while the Hertz potential [Eq.~\eqref{eq:Hertz_pair_potential}] is a two-body energy.

Although $\tilde F_\text{gel}$ is in general a many-body property, in the case of isolated pairs of faceting microgels we may define an effective pair potential $v_{\rm fac}(r)$, which characterizes the free energy cost of forming a single facet. For two microgels of equal size, whose centers are separated by a distance $r$,
\begin{equation}\label{eq:facet_pair_potential}
v_{\rm fac}(r) = 2\left[\tilde F_{\rm gel}(\tilde{\phi}_p) - F_{\rm gel}(\phi_p)\right],
\end{equation}
where $F_{\rm gel}(\phi_p)$ and $\tilde F_{\rm gel}(\tilde{\phi}_p)$ are the Flory-Rehner free energies of a microgel before [Eq.~\eqref{eq:flory-rehner-equation}] and after [Eq.~\eqref{modified_fr_energy}] faceting, respectively, and $\tilde{\phi}_p$ depends implicitly on $r$. The factor of two reflects the symmetric contribution from both microgels. 

We emphasize that our facet model, as here formulated, applies only to the idealized case of homogeneously 
cross-linked microgels with uniform distribution of polymer chains. However, the model may be extended to 
inhomogeneous (e.g., core-shell) microgel architectures by allowing only the outer shell to facet, while 
keeping the inner core spherical. Another conceivable mechanism by which microgels may respond to crowding
is interpenetration of their polymer networks. To assess the relative favorability of such a response 
for homogeneous microgels, we next consider an alternative model.

\subsection{Penetration Model of Microgels}\label{penetration_model_section}

\begin{figure}[h]
    \centering
    \includegraphics[width=0.9\linewidth]{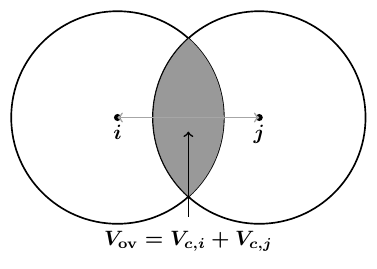}
\caption{
Schematic drawing of two interpenetrating microgels, \(i\) and \(j\). The shaded overlap region, in which the polymer networks of the two microgels interpenetrate, has a volume \(V_{\mathrm{ov}} = V_{c,i} + V_{c,j}\). Each microgel contributes the volume of a spherical cap (\(V_{c,i}\) or \(V_{c,j}\)) to this region. 
}
\label{fig:penetration_model}
\end{figure}

Inspired by the experiments of Conley \textit{et al.}\cite{Conley2017Jamming,conley2019,scheffold2020pathways} 
and simulations of Del Monte and Zaccarelli,\cite{zaccarelli2024} which suggest that microgels in
concentrated suspensions may interpenetrate, we consider, as an alternative to the facet model, a model 
in which microgels respond to crowding by interpenetrating (overlapping), as illustrated in Fig.~\ref{fig:penetration_model}. In the overlap region between two interpenetrating microgels, the respective polymer networks mix, locally increasing the polymer concentration. 
A pair penetration free energy can be defined as the difference in free energy between overlapping and 
non-overlapping microgels. Neglecting possible changes in conformational entropy of the polymer network due to 
crowding and entanglement, the penetration free energy equals the change in polymer-solvent mixing entropy.

Within the Flory-Huggins mean-field approximation,\cite{FloryHugginsMixing} the polymer-solvent mixing free energy associated with an overlap region of volume $V_{\rm ov}$, containing $N_s=V_{\rm ov}\phi_s/v_s$ solvent molecules and a uniform polymer volume fraction $\phi_p$, is given by
\begin{equation}\label{fmixing}
\beta F_{\text{mix}} 
= \frac{V_{\rm ov}}{v_s}(1-\phi_p)\Big[\ln(1-\phi_p)+\chi\phi_p\Big],
\end{equation}
where the entropy of the high molecular weight polymer network is negligible.
For two interpenetrating microgels, $i$ and $j$, of radii $a_i$ and $a_j$ 
and polymer volume fractions $\phi_{pi}$ and $\phi_{pj}$, 
whose centers are separated by a distance $r$, and whose polymers overlap in spherical cap-shaped regions of volumes $V_{c,i}$ and $V_{c,j}$, the lens-shaped overlap volume, $V_{\rm ov}=V_{c,i}+V_{c,j}$, is given by
\begin{equation}\label{Vov}
V_{\rm ov} = \frac{\pi}{12r}(a_i+a_j-r)^2[r^2+2r(a_i+a_j)+6a_ia_j-3(a_i^2+a_j^2)].
\end{equation}

Because this free energy is a property of only two microgels and depends only on the particles' separation, it can be interpreted as an effective pair potential:
\begin{equation}\label{pair_penetration_energy2}
v_{\mathrm{pen}}(r) = \frac{V_{\rm ov}(r)}{v_s}
\Big[f(\phi_{pi}+\phi_{pj}) - f(\phi_{pi}) - f(\phi_{pj})\Big],
\end{equation}
with $\beta f(\phi) \equiv (1-\phi)[\ln(1-\phi) + \chi\phi]$.
Thus defined, $v_{\mathrm{pen}}(r)$ involves only the change in mixing free energy upon penetration. 
The total free energy, including both single-particle swelling and pair interpenetration free energies, then can be expressed as
\begin{equation}\label{total_free_energy_penetration_model}
F_{\text{pen}} = \sum_{i=1}^{N_m} F_{\text{gel}}(\alpha_i) + U_{\rm pen},
\end{equation}
where 
\begin{equation}\label{total_penetration_energy}
U_{\text{pen}} = \sum_{i<j=1}^{N_m} v_{\text{pen}}(r_{ij}; \alpha_i, \alpha_j)
\end{equation}
is the total interpenetration effective pair (free) energy. Note that the Hertz elastic potential energy 
does not enter this model, having been replaced by a penetration free energy. 

As shown below in Sec.~\ref{results}, for typical system parameters, $v_{\mathrm{pen}}(r)$ proves to be
much more strongly repulsive than either the Hertz potential or the facet model effective pair potential. 
As a result, homogeneous microgels whose response to contact is limited to penetration are driven to 
significantly deswell in crowded suspensions, which likely is physically unrealistic. We emphasize, 
however, that our penetration model, although introduced here for the idealized case of homogeneous 
polymer and cross-link densities, may be adapted to microgels with nonuniform (e.g., core-shell) 
internal structure. As discussed further in Sec.~\ref{swelling_ratio_section}, our model could then 
be applied to initial penetration of loosely cross-linked outer shells, followed possibly by 
compression of densely cross-linked inner cores, should they ever make contact.

\section{Computational Methods}\label{methods}

\subsection{Monte Carlo Simulations}\label{MC}

\begin{figure}[h]
  \centering
  \includegraphics[width=0.9\linewidth]{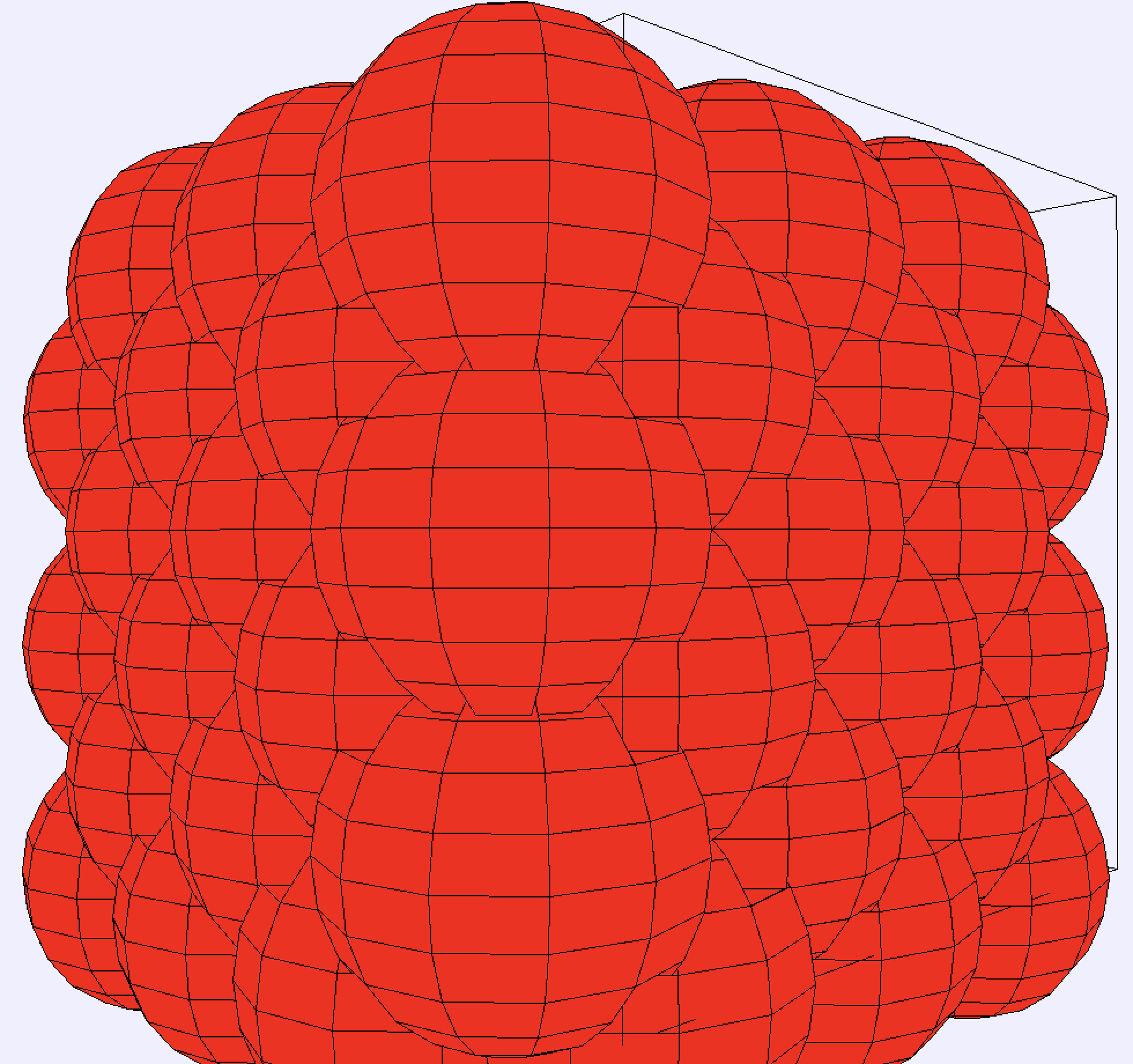}
  \caption{
Snapshot from a simulation of $N_m=108$ microgels (red spheres) within a cubic simulation box in the 
canonical ensemble. Particles in their instantaneous swollen states are shown faceting or interpenetrating.}
  \label{fig:snapshot}
\end{figure}

To investigate the phase behavior of suspensions of microgels that swell and interact according to the facet and penetration models described in Sec.~\ref{models}, we performed Monte Carlo simulations of $N_m$ compressible microgels confined to a cubic box of volume $V$ in the canonical ensemble. To minimize finite-size effects, periodic boundary conditions were applied to all faces of the box. Trial moves consisted of displacements of particle centers (${\bf r}\to {\bf r}+\Delta{\bf r}$) and changes in particle radii ($a \to a+\Delta a$). The trial changes, $\Delta{\bf r}$ and $\Delta a$, were randomly generated within tolerances chosen to optimize accuracy and computational efficiency. The particles were initialized on the sites of a face-centered cubic (FCC) lattice and assigned initial radii equal to the fully swollen radius. The FCC lattice, being the most efficient packing of spheres, ensures that particles initially are maximally separated. Furthermore, FCC is the equilibrium crystal structure into which a fluid of Hertzian spheres freezes.\cite{Pamies2009Hertzian} We custom-coded all simulations in Java within the Open Source Physics Library.\cite{osp-sip2007,osp2006} Figure~\ref{fig:snapshot} shows a typical simulation snapshot.

A simulation proceeds by a sequence of MC steps, each involving a trial move performed on every particle, each move being accepted with a probability following the Metropolis algorithm.\cite{BinderHeermann2010,FrenkelSmit2023} In the facet model, the acceptance probability is given by
\begin{equation}\label{Pacc_facet}
    P_{\mathrm{acc}} = \min\left\{1, \exp\left[-\beta(\Delta\tilde F_{\rm gel}+\Delta U_H)\right]\right\},
\end{equation}
where $\Delta\tilde F_{\rm gel}$ and $\Delta U_H$ are changes in the total modified Flory-Rehner gel free energy [Eq.~\eqref{modified_fr_energy}] and total Hertz pair interaction energy [Eq.~\eqref{eq:internal_energy}], respectively. In the penetration model, the acceptance probability is given by
\begin{equation}\label{Pacc_pen}
    P_{\mathrm{acc}} = \min\left\{1, \exp\left[-\beta(\Delta F_{\rm gel}+\Delta U_{\rm pen})\right]\right\},
\end{equation}
where $\Delta F_{\rm gel}$ and $\Delta U_{\rm pen}$ are changes in the total Flory-Rehner gel free energy [Eq.~\eqref{eq:flory-rehner-equation}] and total interpenetration effective pair energy 
[Eq.~\eqref{total_penetration_energy}], respectively. Accepted moves result in updating of particle positions and radii, while rejected moves leave the state unchanged. During a simulation, as particles move and change size, we keep track of cap volumes associated with facets and overlaps to update all gel free energies and pair energies.

\subsection{Lindemann Criterion for Melting of Solids}\label{Lindemann_Ratio}

Melting of crystalline solids can be diagnosed via the Lindemann criterion,\cite{Lindemann1910} according to which a solid becomes unstable when the root-mean-square displacement of particles from their equilibrium lattice sites $\langle|{\bf r}-{\bf r}_0|^2\rangle^{1/2}$ exceeds a critical fraction of the mean nearest-neighbor distance $d_{nn}$. Here ${\bf r}_0$ are lattice positions and angular brackets denote an ensemble average over configurations. This criterion has been applied to hard-sphere crystals,\cite{HooverRee1968} molecular solids,\cite{Hoffmann2004OnLM} and more recently to colloids and low-dimensional systems,\cite{Khrapak2020} providing a simple yet effective indicator of solid–fluid transitions.
Since the criterion depends only on the particles' center-of-mass coordinates, and not on their
internal structure, it applies equally well to systems of particles with internal degrees of freedom 
(e.g., fluctuating particle radii).

The Lindemann ratio, defined as
\begin{equation}\label{L}
    L = \frac{\langle|{\bf r}-{\bf r}_0|^2\rangle^{1/2}}{d_{nn}},
\end{equation}
typically lies in the range \(0.1 \leq L \leq 0.2\) at the melting point of most materials.\cite{Hoffmann2004OnLM} In our simulations, we evaluated $L$ from Eq.~\eqref{L} by computing the displacement of every particle from its initial lattice site and averaging over particles and MC steps. We corrected for drift of the system's center-of-mass (CM) by applying a constant shift to every particle at each step to zero the CM. As the melting criterion for our suspensions, we chose a value of $L=0.15$, a threshold below which microgel displacements began to be significantly reduced.

We found the Lindemann melting criterion to be robust, with $L$ varying rapidly near the fluid-solid 
transition, making the computed phase stability boundary insensitive to small variations of the threshold 
value (see Sec.~\ref{lindemann_ratio_section}). Increasing the concentration inhibits particle 
mobility and promotes crystalline order, as reflected by lower values of $L$. Conversely, decreasing 
the concentration facilitates particle mobility, which can destabilize a crystal and induce melting,
as heralded by higher values of $L$. As further discussed in Sec.~\ref{structure_section}, we 
independently corroborated phase identification by computing structural properties of the system. 
For states identified as being in a crystalline state, the radial distribution function and static 
structure factor exhibit sharp peaks characteristic of long-range ordering.

\section{Results and Discussion}\label{results}

\subsection{Effective Pair Potentials}\label{effective_pair_potentials}

As discussed in Sec.~\ref{models}, microgels can interact with one another by interpenetrating their polymer networks, by deforming to create facets, or by some combination of these mechanisms. The interaction energy between two microgels can be quantified by an effective pair potential $v_{\rm eff}(r)$, which provides insight into the influence of single-particle properties (e.g., cross-link fraction) and solvent conditions on bulk phase behavior. In the Hertz theory, elastic deformation energy of interacting microgels is described by the Hertz potential $v_H(r)$ [Eq.~\eqref{eq:Hertz_pair_potential}]. In the penetration model, we defined an effective pair potential $v_{\rm pen}(r)$ associated with free energy of penetration [Eq.~\eqref{pair_penetration_energy2}]. Similarly, in the facet model, we defined an effective pair potential $v_{\rm fac}(r)$ associated with redistribution of polymer to accommodate a facet [Eq.~\eqref{eq:facet_pair_potential}], noting, however, that the free energy then is a many-body property that may not be expressible simply as a sum of pair potentials.
 
\begin{figure}[h]
\centering
\includegraphics[width=1.0\linewidth]{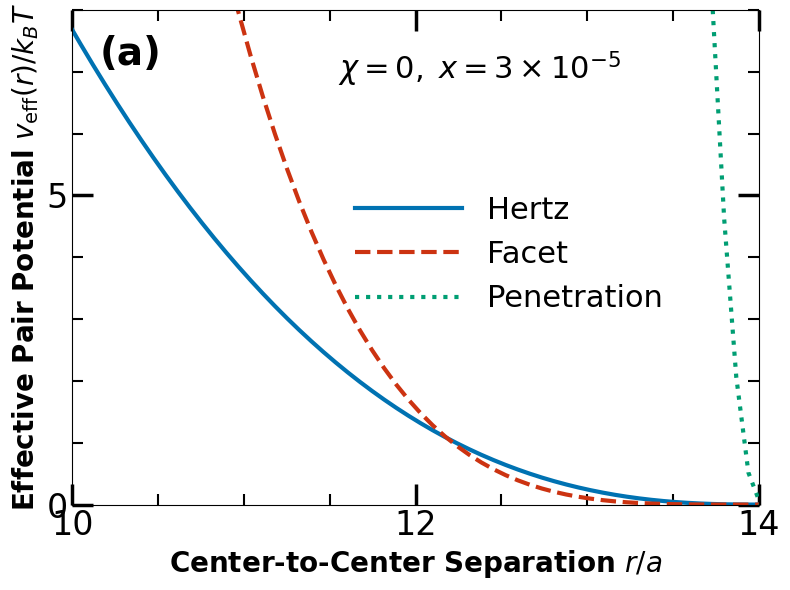} \\[2ex]
\includegraphics[width=1.0\linewidth]{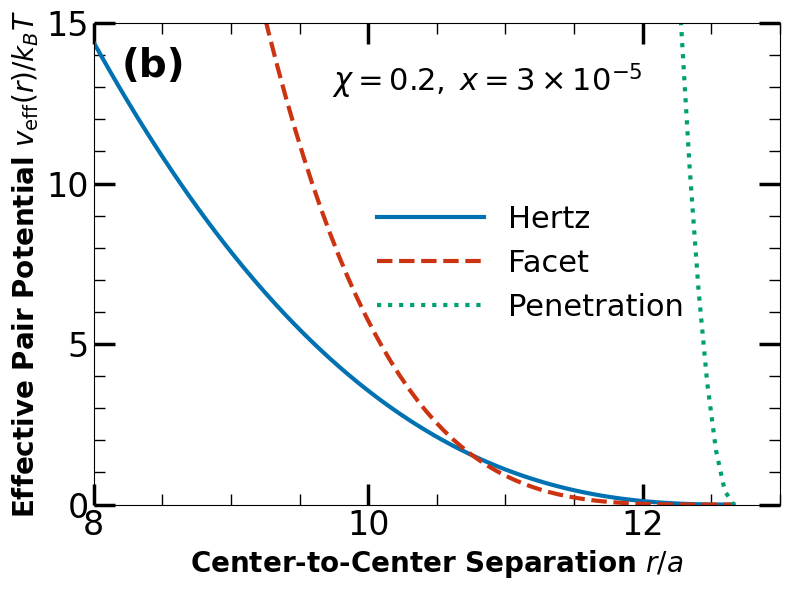}
\caption{Effective pair potentials between microgels dispersed in solvents of differing qualities:
(a) $\chi=0$ (good) and (b) $\chi=0.2$ (moderately good). 
Results are shown for three models: Hertz (blue solid curves) [Eq.~\eqref{eq:Hertz_pair_potential}], 
facet (red dashed curves) [Eq.~\eqref{eq:facet_pair_potential}],
and penetration (green dotted curves) [Eq.~\eqref{pair_penetration_energy2}].
Other system parameters are as in Sec.~\ref{effective_pair_potentials}.}
\label{fig:effective_pair_potential}
\end{figure}

We numerically evaluated $v_{\mathrm{eff}}(r)$ within the different models for microgels with dry radius $a_0=50$ nm, polymer monomer radius $r_{\mathrm{mon}}=0.3$ nm, cross-link fraction $x=3\times 10^{-5}$, and Young’s calibration factor $Y_c=1$, dispersed in solvents with $\chi=0$ and $\chi=0.2$, corresponding to good and moderately good solvent conditions. As shown in Fig.~\ref{fig:effective_pair_potential}, the effective pair potential varies strongly between the models. In the penetration model, $v_{\mathrm{eff}}(r)$ rises steeply near contact, reflecting the relatively high entropic cost of network overlap. 
We emphasize, however, that our penetration model, at least in its present form, treats the microgels 
as having uniformly distributed polymer segments and cross-links, whereas most real microgels may have a 
diffuse and more loosely cross-linked corona (shell) surrounding a denser core. Our results, therefore, 
may not accurately describe the interactions of neighboring shells, but may rather demonstrate the 
tendency for inner cores to strongly resist penetration.

In contrast, the facet model exhibits a much softer repulsive potential at short separations, reflecting the moderate entropic cost of polymer redistribution. The Hertz potential is similarly soft, capturing the energetic cost of elastic deformation, but slightly more repulsive than the facet potential near contact. These results provide the first indication that pair energy considerations may favor faceting over interpenetrating for 
uniformly cross-linked microgels with internal freedom to respond to crowding.

Also seen in Fig.~\ref{fig:effective_pair_potential} is the influence of solvent quality on $v_{\mathrm{eff}}(r)$. For $\chi=0$ (good solvent), strong polymer-solvent affinity promotes swelling and yields a larger equilibrium swelling ratio $\alpha$. For $\chi=0.2$, poorer solvent quality favors polymer-polymer interactions, reducing the osmotic driving force for swelling and leading to more compact particles that more strongly repel.

Evidently, and interestingly, the Hertz potential [Eq.~(\ref{eq:Hertz_pair_potential})] and the facet model 
effective pair potential [Eq.~(\ref{eq:facet_pair_potential})] can have comparable magnitudes for some 
system parameters. Furthermore, they are both softly repulsive and naturally have the same range 
(sum of particle radii). It is perhaps not surprising, then, that the Hertz potential proves qualitatively 
reasonable in describing microgel pair interactions. However, $v_{\rm fac}(r)$ scales differently with 
compression depth $h$ for $h\ll a$ ($\sim h^2$ or $h^4$, depending on system parameters). 
Moreover, while Hertzian interactions are strictly pairwise, the facet model accounts for 
many-body interactions.

As discussed in Sec.~\ref{facet_model_section}, whether the facet model swelling free energy should 
supplement or replace the Hertz pair potential is an interesting and important question. Although we defer 
further discussion to a future study, for comparison, we present results in Sec.~\ref{swelling_ratio_section}
for equilibrium swelling ratios both with and without the Hertz potential included. The latter case may 
correspond to ultrasoft microgels that facet only, with no Hertzian interactions.

\subsection{Equilibrium Lindemann Ratio}\label{lindemann_ratio_section}

To investigate the melting behavior of suspensions of compressible microgels, we performed a series of Monte Carlo simulations of $N_m=4\times 3^3=108$ particles, commensurate with the FCC lattice, within both the penetration and facet models over a range of concentrations at three cross-link fractions: $x=3\times 10^{-5}, 4\times 10^{-5}$, and $5\times 10^{-5}$. 
These values of $x$ represent microgels with softness in the range relevant to the fluid-solid phase diagram 
determined by P{\`a}mies \textit{et al.}\cite{Pamies2009Hertzian} (see Sec.~\ref{stability_curve_section}).
We confirmed that finite-size effects are negligible through test runs with fewer ($32$) and more ($256$) particles. Other system parameters were microgel dry radius $a_0=50$ nm, polymer monomer radius $r_{\mathrm{mon}}=0.3$ nm, Young’s calibration factor $Y_c=1$, and solvency parameter $\chi=0$. In each simulation, the particles were initialized on the sites of an FCC lattice and evolved through trial displacements and size changes, as detailed in Sec.~\ref{MC}. Following an equilibration period of $10^4$ MC steps, we sampled data every 100 steps over an additional $10^5$ steps and computed the equilibrium Lindemann ratio by averaging over configurations. During the simulations, we computed the total energy and pressure, ensuring that after equilibration these properties fluctuated around steady values,
and equilibrium values for the Lindemann ratio and swelling ratio. 

\begin{figure}[h]
    \centering
    \includegraphics[width=0.9\linewidth]{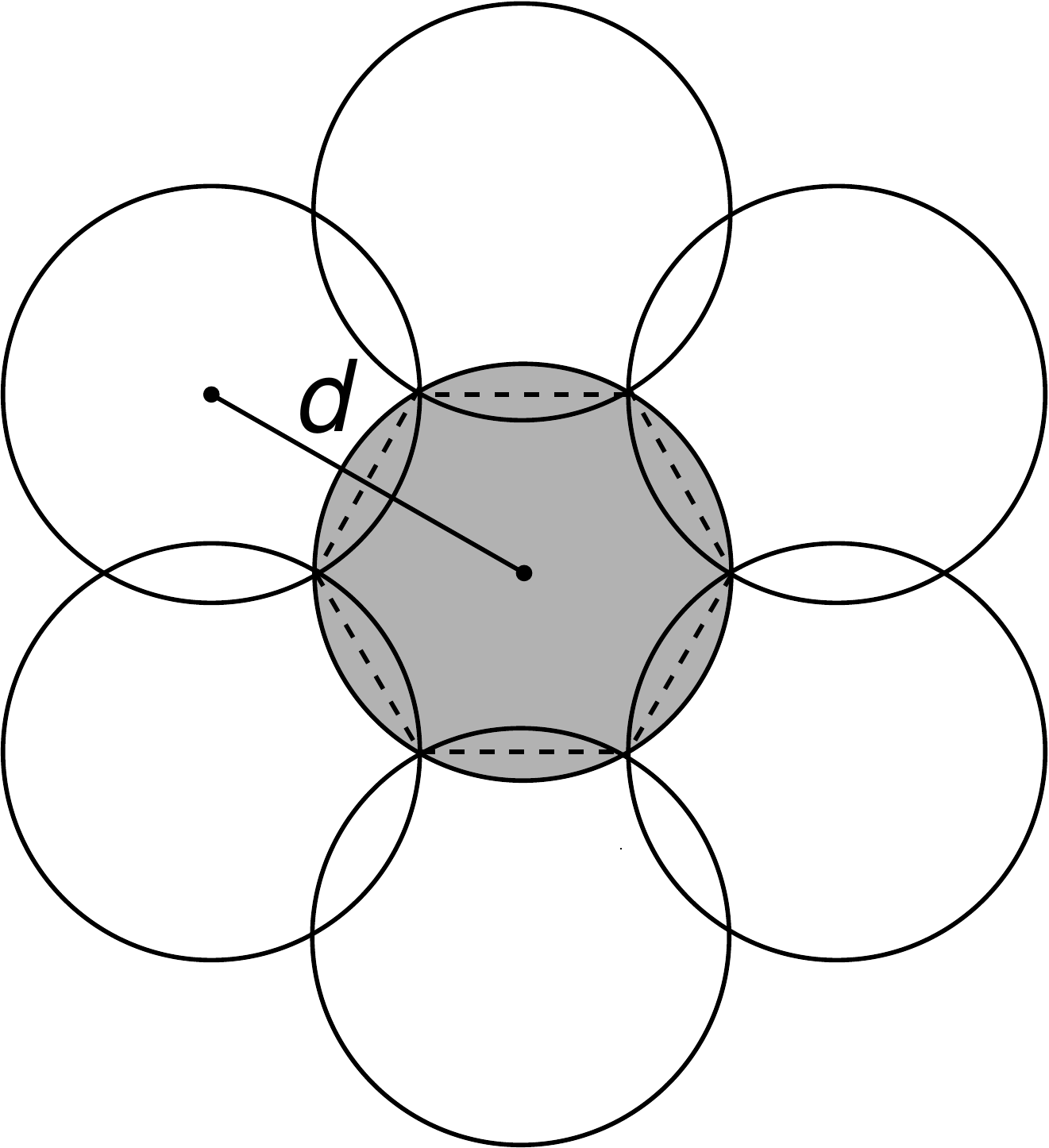}
    \caption{
Schematic drawing of a central microgel (gray) surrounded by neighbors (white) in a crystalline structure. 
This 2D configuration depicts the threshold packing for intersection of neighboring facets (dashed lines)
on the same particle. In our simulations of the restricted facet model in 3D (Sec.~\ref{model_validation}),
with particle swelling, but no displacements, the center-to-center distance $d$ is reduced until 
particles mutually facet.} 
\label{microgel_in_an_fcc_lattice}
\end{figure}

At fixed temperature, phase behavior of microgel suspensions can be controlled by varying the concentration, 
which is proportional to the dry volume fraction, $\phi_0=V_0/V$, where $V_0=(4\pi/3)a_0^3N_m$ is the 
total volume of microgels in their dry state. 
The dry volume fraction is directly proportional to the 
generalized (or nominal) volume fraction, $\zeta=(a_{\rm res}/a_0)^3\phi_0$, defined as the ratio of the 
volume of the microgels in their fully swollen state (with reservoir radius $a_{\rm res}$) to the total volume. 
In contrast, the actual volume fraction $\phi$ is the fraction of volume occupied by microgels with 
swollen (possibly compressed) radii $a_i$ ($i=1,\ldots, N_m$), accounting for faceting by subtracting 
the volumes $V_{c,i}$ of any spherical caps formed by facets:
\begin{equation}\label{phi}
\phi = \frac{1}{V}\sum_{i=1}^{N_m}\left(\frac{4\pi}{3}a_i^3-V_{c,i}\right).
\end{equation}
Thus, $\phi\leq 1$ and $\phi\leq\zeta$, whereas $\zeta$ can significantly exceed 1.

To avoid complications resulting from intersecting facets on the same particle, we maintained 
the swollen volume fraction below the limit at which neighboring facets start to intersect. 
We estimated this limit by considering a perfect FCC crystal at generalized volume fraction 
$\zeta>\zeta_{cp}$, where $\zeta_{cp}=\sqrt{2}(\pi/6)$ is the close-packed volume fraction of
hard spheres, in which case each particle has 12 symmetrically distributed circular facets. 
The diameter of each facet subtends an angle of $60^{\circ}$ at the center. For spheres all of 
unit radius, when neighboring facets on the same sphere are just touching, the nearest-neighbor 
center-to-center distance is $d_{nn}=\sqrt{3}$ and the generalized volume fraction is 
$\zeta=\zeta_{cp}/\cos^3 30^{\circ}=1.1400$. Correcting for the
volume $V_c$ of the 12 spherical caps, the volume fraction in this state is $\phi=\zeta(1-V_c/V_m)$,
where $V_m=4\pi/3$ is the microgel volume. For 12 spherical caps of height $h=1-\cos 30^{\circ}$, 
we have $V_c/V_m=3h^2(3-h)$,
from which we calculate the actual volume fraction $\phi=0.9641$. As shown below in 
Sec.~\ref{stability_curve_section}, this upper limit for applicability of our facet model is 
well above the volume fraction at which the fluid-solid phase transition occurs.
Moreover, microgel deswelling lowers $\phi$ and thus will considerably increase the upper limit of $\zeta$.
For comparison, in two dimensions, a hexagonal crystal has a limiting generalized area fraction 
$\zeta=1.2092$ and, with six facets per circular particle, an actual area fraction $\phi=1$ when 
neighboring facets begin to intersect, as illustrated in Fig.~\ref{microgel_in_an_fcc_lattice}.

\begin{figure}[h]
    \centering
    \includegraphics[width=1.0\linewidth]{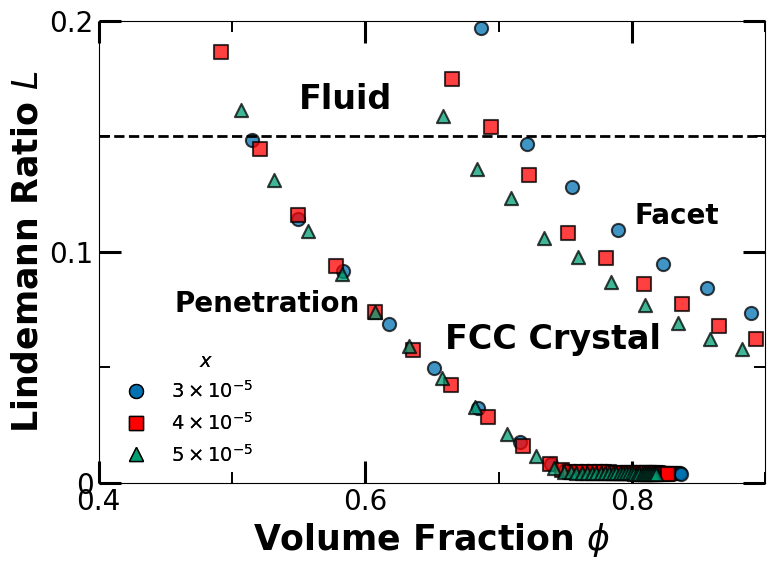}
    \caption{
Simulation data for equilibrium Lindemann ratio $L$ vs.~volume fraction $\phi$ in penetration and facet models of microgels with cross-link fractions $x=3\times 10^{-5}$ (blue circles), $4\times 10^{-5}$ (red squares), and $5\times 10^{-5}$ (green triangles). Other system parameters are as in Sec.~\ref{lindemann_ratio_section}. The horizontal dashed line at $L=0.15$ marks the typical melting threshold separating fluid and FCC crystalline phases.
    }
    \label{lindemann_ratio}
\end{figure}

As shown in Fig.~\ref{lindemann_ratio}, the Lindemann ratio falls below the threshold of $L=0.15$ for a stable (or metastable) solid at significantly different volume fractions in the two models. 
For reference, as the actual volume fraction $\phi$ varies from 0.5 to 0.8, the dry volume fraction $\phi_0$ 
varies from roughly 0.002 to roughly 0.003 (facet model) or 0.004 (penetration model).
We emphasize that the rapid change of $L$ with respect to $\phi$ near the melting transition ensures that
the chosen threshold is robust with respect to minor variations.
In the penetration model, the fluid-solid transition occurs at $\phi\simeq 0.52$, which is nearly the same as for the hard-sphere fluid, and independent of cross-link fraction. This observation suggests that particle softness plays a minimal role in phase behavior within this model. In contrast, in the facet model, the melting transition occurs at considerably higher volume fractions, $\phi\simeq 0.65-0.7$, and exhibits a noticeable dependence on cross-linking. The volume fraction at which the suspension crystallizes steadily increases with increasing particle softness (i.e., decreasing $x$). This behavior is consistent with the experimental observations of Pelaez-Fernandez \etalia\cite{PelaezFernandez2015} (discussed in Sec.~\ref{introduction}), who found that microgel suspensions crystallized at higher volume fractions the softer the particles. This qualitative agreement with experiment reinforces the conclusion that compressibility of microgels strongly influences their phase behavior. In Sec.~\ref{stability_curve_section}, we present a more in-depth analysis of phase stability.

\begin{figure}[h]
    \centering
    \includegraphics[width=1.0\linewidth]{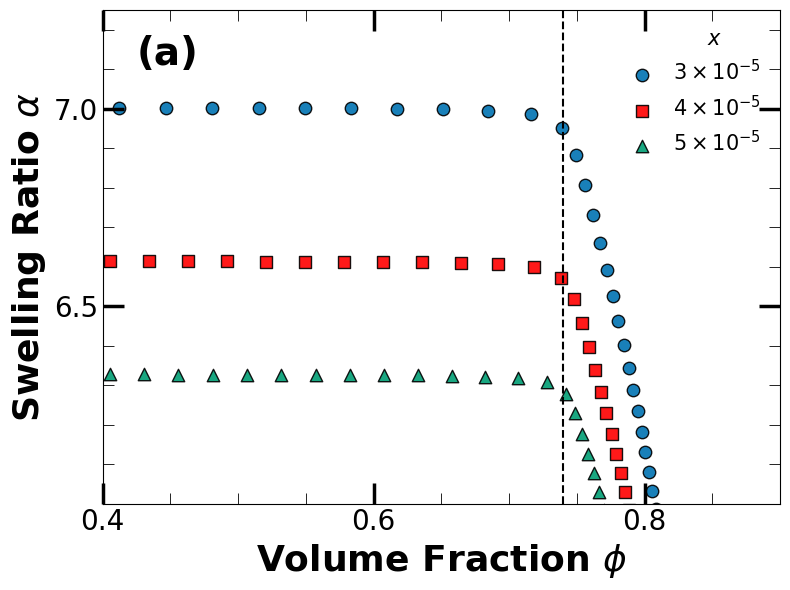}
    \includegraphics[width=1.0\linewidth]{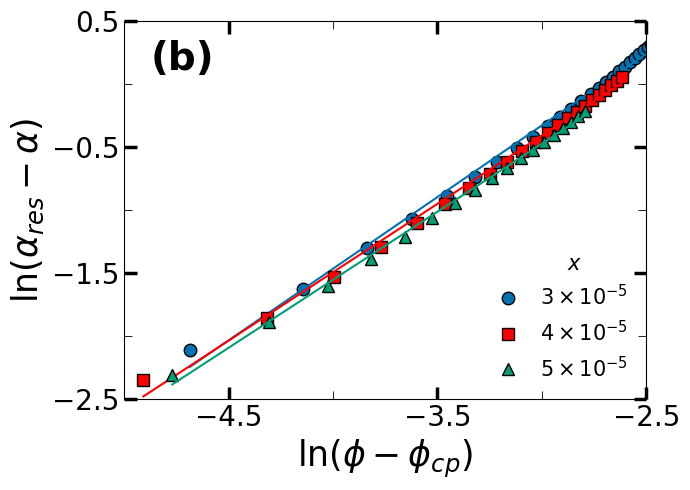}
    \caption{
(a) Simulation data for equilibrium swelling ratio $\alpha$ vs.~volume fraction $\phi$ in the penetration model of microgels with cross‐link fractions $x=3\times 10^{-5}$ (blue circles), $4\times 10^{-5}$ (red squares), and $5\times 10^{-5}$ (green triangles). Other system parameters are as in Sec.~\ref{lindemann_ratio_section}. Increasing $x$ stiffens the polymer network and lowers the equilibrium swelling ratio. The vertical dashed line marks the hard‐sphere close‐packing limit $\phi_{cp}\simeq 0.74$, beyond which microgels exhibit rapid deswelling with increasing $\phi$. (b) Same data on a log-log scale. Straight lines are least-squares fits to $\alpha_{\rm res}-\alpha\sim(\phi-\phi_{cp})^{\gamma}$, demonstrating power-law scaling for $\phi>\phi_{cp}$.}
    \label{swelling_ratio_pen}
\end{figure}

\begin{figure}[h]
    \centering
    \includegraphics[width=1.0\linewidth]{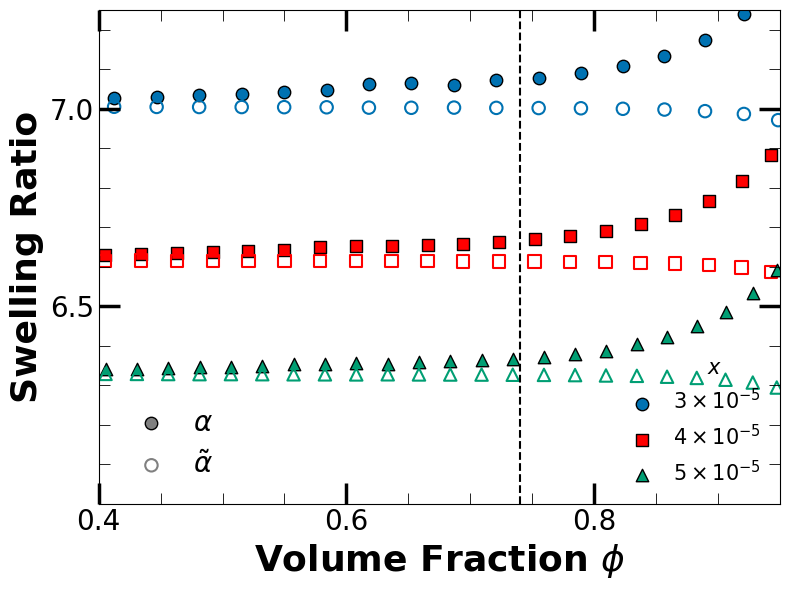}
    \caption{
Simulation data for equilibrium swelling ratio vs.~volume fraction $\phi$ in the facet model of microgels with cross‐link fractions $x=3\times 10^{-5}$ (blue circles), $4\times 10^{-5}$ (red squares), and $5\times 10^{-5}$ (green triangles). Other system parameters are as in Sec.~\ref{lindemann_ratio_section}. Increasing $x$ stiffens the polymer network and lowers the equilibrium swelling ratio. Solid and open symbols represent, respectively, the swelling ratio $\alpha$ and modified swelling ratio $\tilde\alpha$, the latter being relatively flat. The vertical dashed line marks the hard‐sphere close‐packing limit $\phi_{cp}\simeq 0.74$, beyond which $\alpha$ gradually increases with increasing $\phi$.}
    \label{swelling_ratio_facet}
\end{figure}

\subsection{Equilibrium Swelling Ratio}\label{swelling_ratio_section}

From the same simulations used to compute the Lindemann ratio of microgel suspensions (Sec.~\ref{lindemann_ratio_section}), we computed the equilibrium swelling ratio $\alpha$ over ranges of volume fraction $\phi$ and cross‐link fraction $x$ within the penetration and facet models. As seen in Figs.~\ref{swelling_ratio_pen} and \ref{swelling_ratio_facet}, the swelling ratio in both models depends sensitively on cross-link fraction. The softer the particles (lower $x$) the higher the swelling ratio, owing to weaker elastic restoring forces to resist solvent absorption.\cite{FloryRehner1943I,FloryRehner1943II,FernandezNieves2003} The response to variation in concentration and mutual crowding, however, is qualitatively different between the two models.

In the penetration model [Fig.~\ref{swelling_ratio_pen} (a)], with increasing volume fraction, the microgels deswell, at first very gradually. As $\phi$ approaches the close-packing limit of hard spheres ($\phi_\text{cp}\simeq 0.74$), where neighboring particles begin to make contact, $\alpha$ falls off sharply with increasing $\phi>\phi_\text{cp}$. 
This strong response to crowding is consistent with the relatively high free energy cost for homogeneous
microgels to interpenetrate, as discussed in Sec.~\ref{effective_pair_potentials}. In the penetration 
free energy [Eq.~(\ref{total_free_energy_penetration_model})], the gel contribution $F_\text{gel}$ determines 
the swelling ratio up to close-packing, while the effective pair (free) energy $U_\text{pen}(r)$, 
which accounts for the change in polymer-solvent mixing free energy upon overlap, dominates the 
response for $\phi>\phi_\text{cp}$. 

Analysis of our simulation data in this regime reveals that the swelling ratio follows a power-law scaling,
\begin{equation}
\alpha_{\rm res}-\alpha\sim(\phi-\phi_{cp})^{\gamma} \quad (\phi>\phi_{cp}),
\label{scaling}
\end{equation}
where $\alpha_{\rm res}$ is the fully swollen (reservoir) swelling ratio. As demonstrated in 
Fig.~\ref{swelling_ratio_pen}(b), the data fall on straight lines on a log-log plot, with fitted exponents 
$\gamma\simeq 1.17, 1.12$, and 1.12 for $x=3\times 10^{-5}, 4\times 10^{-5}$, and $5\times 10^{-5}$,
respectively, indicating near-linear scaling and independence of cross-link fraction. 

We emphasize that the current incarnation of our penetration model for idealized homogeneous microgels
probably does not accurately describe the response to contact of microgels with core-shell architecture,
whose inner cores are coated by loosely cross-linked coronas and dangling polymer chains. In fact, 
neither experiments~\cite{Conley2017Jamming} nor molecular-scale simulations~\cite{zaccarelli2024} 
of core-shell microgels appear to show any evidence of penetration of the inner cores. Thus, the 
predicted extreme response to penetration here may rather support the conclusion that real microgels tend 
not to be homogeneous. However, the model could be adapted to describe interactions between only the
diffuse peripheries of microgels before any core contact.

Next, we examine how microgels in the facet model respond to mutual crowding. As shown in 
Fig.~\ref{swelling_ratio_facet}, with increasing volume fraction, loosely cross-linked microgels 
($x\le 5\times 10^{-5}$) 
respond by swelling, gradually at low $\phi$, but more rapidly for $\phi>\phi_\text{cp}$.
Thus, while polymer mixing in the penetration model favors deswelling, faceting favors swelling
of relatively soft microgels, especially in highly crowded environments. At higher cross-link fractions,
however, faceting can alternatively drive deswelling of stiffer microgels (see below).
The predicted swelling behavior is driven by competition between swelling free energy and Hertzian energy 
in minimizing the total free energy [Eq.~(\ref{F_fac})]. As seen in Fig.~\ref{swelling_ratio_facet}, 
with increasing $\phi$, sufficiently soft microgels with relatively weak Hertzian pair repulsion ($Y_c=1$)
grow in radius ($\alpha$ increases), since the swelling free energy [Eq.~(\ref{nonlocalfacet})] dominates.  
At the same time, the {\it modified} swelling ratio $\tilde\alpha$ [Eq.~(\ref{tilde_alpha1})] remains 
relatively constant, only gradually decreasing, reflecting near conservation of local polymer volume 
fraction $\phi_p$.

\begin{figure}[h]
    \centering
    \includegraphics[width=1.0\linewidth]{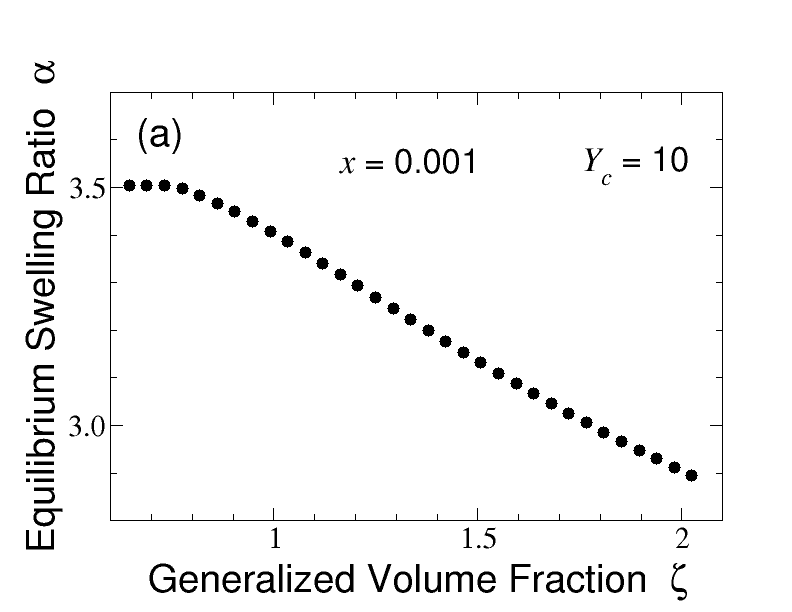}
    \includegraphics[width=1.0\linewidth]{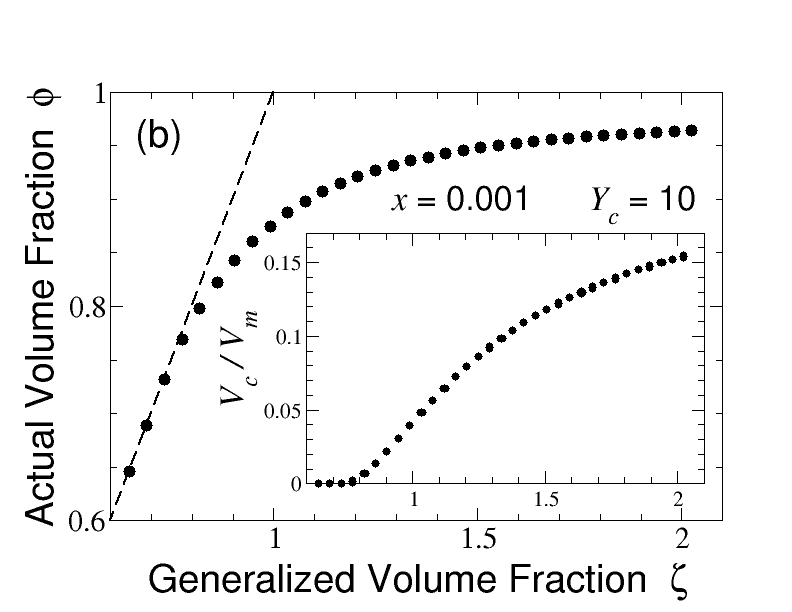}
    \caption{
Simulation data for (a) equilibrium swelling ratio $\alpha$ and (b) mean actual volume fraction $\phi$ 
vs.~generalized volume fraction $\zeta$ in the facet model of microgels with cross‐link fraction 
$x=0.001$ and Young's calibration factor $Y_c=10$.
Other system parameters are as in Sec.~\ref{effective_pair_potentials}. 
Straight dashed line in panel (b) represents $\phi=\zeta$.
Inset: Mean overlap volume fraction $V_c/V_m$.
Beyond the hard‐sphere close‐packing limit of $\zeta\simeq 0.74$, microgels deswell and form facets 
as $\alpha$ gradually decreases, $\phi$ falls below $\zeta$, and $V_c/V_m$ steadily increases 
with increasing $\zeta$.
}
    \label{facet_stiff}
\end{figure}

For comparison, we also explore the response to crowding of somewhat stiffer microgels, characterized by 
$x=0.001$ and $Y_c=10$. As shown in Fig.~\ref{facet_stiff}, with increasing generalized volume fraction, 
these particles initially maintain their swollen size. When $\phi$ exceeds $\phi_\text{cp}$, however, 
the equilibrium swelling ratio starts to monotonically decrease. At the same time, as facets emerge and grow,
the mean overlap volume fraction $V_c/V_m$ gradually increases. Consequently, the mean actual volume fraction 
$\phi$ progressively falls below $\zeta$, as it approaches its maximum value of 1. Recall that our
$\phi$ is defined [Eq.~(\ref{phi})] to correct for pair overlaps of microgels. When $\zeta\simeq 2.02$,
we compute $\phi\simeq 0.964$, beyond which neighboring facets would start to intersect.

These predictions may be compared with corresponding results from the molecular dynamics simulations of 
Del Monte and Zaccarelli.\cite{zaccarelli2024} In particular, their Figs.~2, 4, 6, and 7 demonstrate 
remarkably similar qualitative swelling behavior, despite the differing system parameters and structure 
(core-shell) of their particles compared with our larger, homogeneously structured microgels.
The same level of qualitative agreement is apparent upon comparing our Fig.~\ref{facet_stiff} with 
the experimental observations reported in Figs.~4 and 5 of Conley \textit{et al.},\cite{Conley2017Jamming}
which show compression (deswelling) and deformation (faceting) of microgel cores.

\begin{figure}[h]
    \centering
    \includegraphics[width=0.9\linewidth]{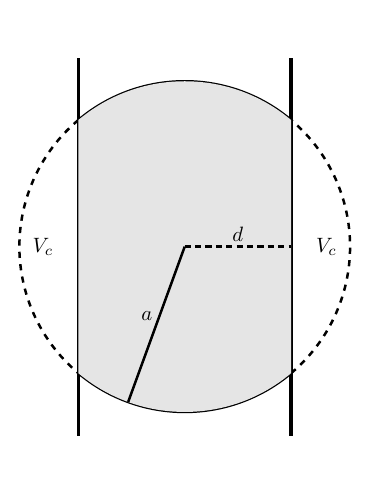}
    \caption{Schematic drawing of a microgel of radius $a$ confined between two planar hard walls separated by distance $2d$, a simplified model for confinement-induced faceting and deswelling in dense suspensions. The shaded region indicates the confined microgel, whose swollen volume is reduced by that of the flattened caps of volume $V_c$.}
    \label{fig:microgel_between_walls}
\end{figure}

\begin{figure}[h]
    \centering
    \includegraphics[width=0.9\linewidth]{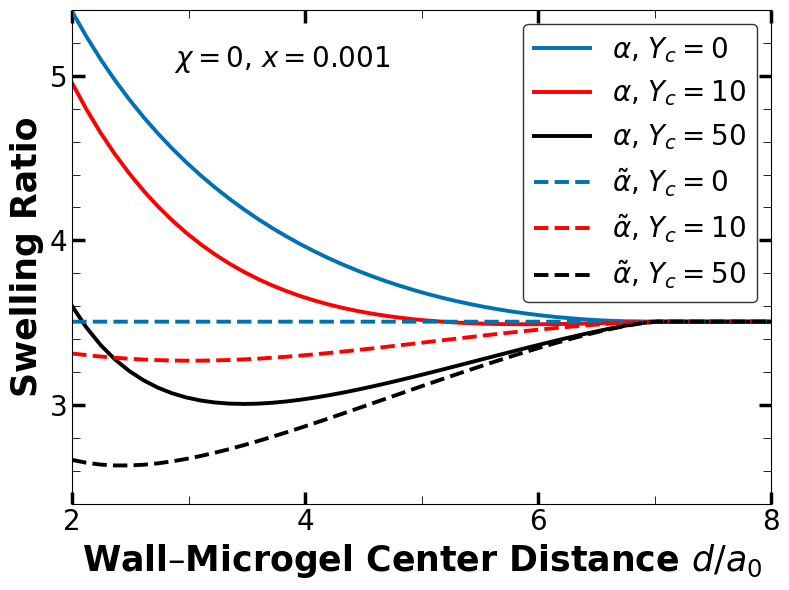}
    \caption{
Equilibrium swelling ratio vs.~wall–microgel-center distance $d/a_0$ for a microgel in the facet model confined between two planar hard walls with cross-link fraction $x=0.001$, solvency $\chi=0$, and Young's calibration factor $Y_c=0$ (no Hertzian interactions, blue curves), 10 (red curves), and 50 (black curves). Other system parameters are as in Sec.~\ref{effective_pair_potentials}. Solid and dashed curves represent, respectively, swelling ratio $\alpha$ and modified swelling ratio $\tilde\alpha$. With decreasing $d$, soft microgels ($Y_c=0$) monotonically swell ($\alpha$ increases), while stiffer microgels ($Y_c=$10, 50) initially deswell and then reswell at narrower wall separations. In all cases, $\tilde\alpha<\alpha$.}
\label{microgel_between_walls}
\end{figure}

To further probe the swelling response of microgels to faceting, we modeled a single microgel placed between two parallel hard walls (see Fig.~\ref{fig:microgel_between_walls}). When not in contact with a wall, the microgel maintains its equilibrium swollen volume. When compressed by the walls, however, faceting removes two caps of volume $V_c$ [Eqs.~\eqref{Vcij} and \eqref{hij}], yielding a modified swelling ratio
\begin{equation}\label{tilde_alpha2}
\tilde{\alpha} = \alpha \left(1-\frac{2V_c}{V_m} \right)^{1/3}.
\end{equation}
This modified swelling ratio replaces $\alpha$ in the Flory–Rehner free energy [Eq.~\eqref{eq:flory-rehner-equation}] to account for the volume lost to confinement.

The Hertz potential between a microgel and a hard wall can be derived from the potential between two spherical microgels [Eq.~\eqref{eq:Hertz_pair_potential}] in the limit as one of the spheres grows to infinite size ($a_j\to\infty$), leading to a microgel-wall interaction energy~\cite{LandauLifshitzElasticity1986} 
\begin{equation}\label{vHd}
    v_H(d) = Y_cN_{\text{ch}} \left(1 - \frac{d}{a}\right)^{5/2},
\end{equation}
where $d$ is the distance between each wall and the microgel center (half the wall separation). Allowing the microgel to respond to confinement by both faceting and swelling/deswelling, we minimized the sum of the wall-sphere Hertz potential energy [Eq.~(\ref{vHd})] and the modified Flory–Rehner free energy evaluated with $\tilde{\alpha}$ from Eq.~(\ref{tilde_alpha2}).
With only one microgel to consider, no simulations were required.

For illustration, we performed calculations under good solvent conditions ($\chi=0$) for a microgel with 
dry radius $a_0=50$ nm, monomer radius $r_{\text{mon}}=0.3$ nm, cross‐link fraction $x=0.001$, and 
three different Young's calibration factors, $Y_c=0$, 10, and 50 [see Sec.~\ref{Hertz_model_section}
and Eq.~(\ref{eq:Y_scaling})].
The case $Y_c=0$ corresponds to ultrasoft microgels with no Hertzian pair interactions, while the 
cases $Y_c=10$ and 50 represent stiffer microgels with relatively strong Hertzian interactions.
From the results shown in Fig.~\ref{microgel_between_walls}, we observe that for $d<a$ confinement 
induces monotonic swelling in the softest microgels ($Y_c=0$), while stiffer particles ($Y_c=10, 50$) 
initially deswell with decreasing $d$ before reswelling at narrower separations. These contrasting 
responses reflect competition between swelling free energy and Hertzian energy and highlight the role 
of Hertzian elasticity in confined environments.

The modified swelling ratio $\tilde\alpha$ remains relatively constant, as we observed also for 
suspensions of faceted microgels, becoming independent of wall separation when Hertzian interactions 
are turned off ($Y_c=0$). As discussed in Sec.~\ref{facet_model_section}, we combine the Hertzian and
facet models on the premise that deformation of the polymer networks near the contact region is a
localized Hookean elastic response, while redistribution of polymer and solvent is a nonlocal response,
affecting the entire microgel, with an associated change in polymer-solvent entropy.

\subsection{Validation of Facet Model Methods}\label{model_validation}

To validate our implementation of the facet model, we modeled an FCC crystal arrangement of microgels (see Fig.~\ref{microgel_in_an_fcc_lattice}) that are free to change size but forced to remain on their lattice sites. For this restricted facet model, we performed MC simulations with the displacement tolerance set to zero. With increasing concentration (decreasing center-to-center distance $d$), microgels begin to form facets with neighboring particles. Under these conditions, the modified swelling ratio of the central microgel is
\begin{equation}\label{tilde_alpha3}
\tilde{\alpha} = \alpha\left(1-\frac{V_{c,\text{total}}}{V_m}\right)^{1/3},
\end{equation}
where $V_{c,\text{total}}$ is the sum of cap volumes due to facets with all 12 neighbors. For comparison, we also performed theoretical calculations by minimizing the sum of the sphere-sphere Hertz potential [Eq.~(\ref{eq:Hertz_pair_potential})] and the Flory–Rehner free energy evaluated with $\tilde{\alpha}$ from Eq.~(\ref{tilde_alpha3}). For both simulations and theory, we chose the same system parameters as in Sec.~\ref{lindemann_ratio_section} and considered ultrasoft ($Y_c=0$) and stiffer ($Y_c=10$) microgels.

\begin{figure}[h]
    \centering
    \includegraphics[width=1.0\linewidth]{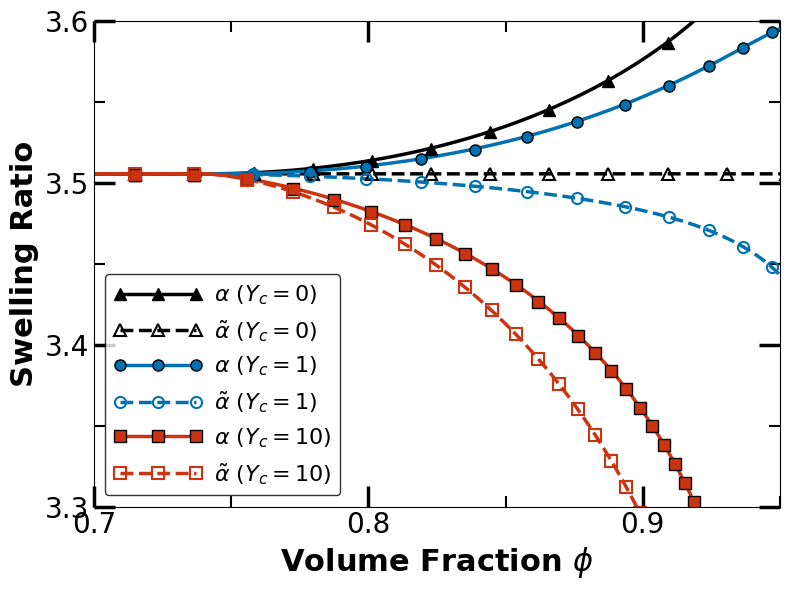}
    \caption{
Equilibrium swelling ratio vs.~volume fraction $\phi$ for microgels in the restricted facet model (particle swelling, but no displacements) with cross-link fraction $x=0.001$, solvency $\chi=0$, and Young's calibration factor $Y_c=0$ (no Hertzian interactions) and $Y_c=10$. Other system parameters are as in Sec.~\ref{effective_pair_potentials}. Solid and open symbols are simulation data, and solid and dashed curves theoretical predictions, for swelling ratio $\alpha$ and modified swelling ratio $\tilde\alpha$, respectively. With increasing $\phi$, soft microgels ($Y_c=0, 1$) swell, while stiffer microgels ($Y_c=10$) deswell. In all cases, $\tilde\alpha<\alpha$.}
\label{facet_model_validation}
\end{figure}

Figure~\ref{facet_model_validation} shows close agreement of swelling ratios calculated from simulations and theory for ultrasoft ($Y_c=0$), soft ($Y_c=1$), and stiffer ($Y_c=10$) microgels in the weak-to-moderate faceting regime. Interestingly, while the soft microgels respond to mutual crowding (increasing $\phi$) by swelling, the stiffer microgels respond by deswelling. 
The latter response is qualitatively comparable to the deswelling observed with increasing $\phi$ beyond 
hard‐sphere close‐packing in the experiments of Pelaez-Fernandez \textit{et al.}\cite{PelaezFernandez2015}
with relatively stiff PVP microgels.
Note that the swelling trend observed here for $Y_c=1$ is similar to that in Fig.~\ref{swelling_ratio_facet} (but for different $x$). Note further that, in the $Y_c=10$ case, the deswelling observed in this FCC confinement, where a microgel has 12 facets, contrasts with the swelling seen in Fig.~\ref{microgel_between_walls} for hard wall confinement, where the microgel has only two facets,
highlighting the relevance of confinement geometry.

\begin{figure}[h]
    \centering
    \includegraphics[width=1.0\linewidth]{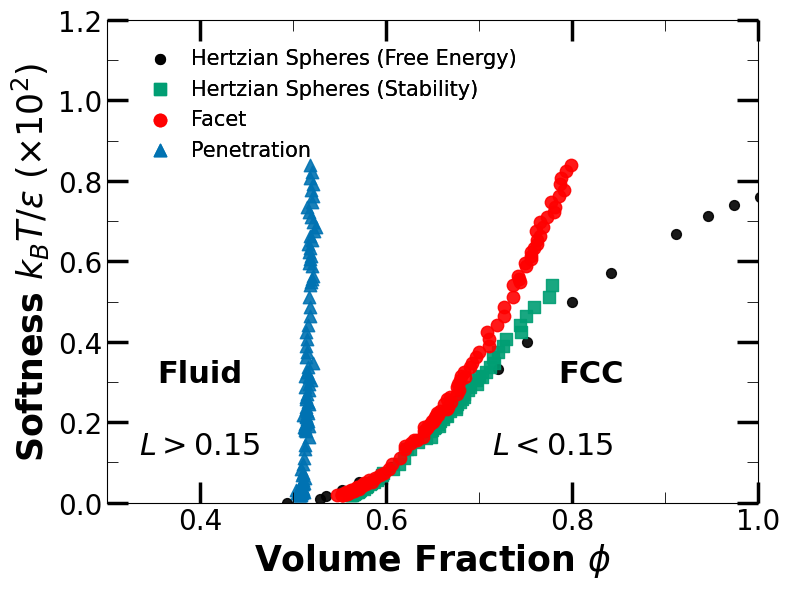}
    \caption{Simulation data for phase stability of suspensions of $N_m=108$ microgels in facet model (red circles), penetration model (blue triangles), and Hertzian-sphere model (green squares). Data show fluid–solid transition, identified using the Lindemann criterion, in the plane of volume fraction $\phi$ and softness $k_B T/\varepsilon$. System parameters are as in Sec.~\ref{effective_pair_potentials}. Also shown is the thermodynamic phase boundary determined by P\`amies \textit{et al.}\cite{Pamies2009Hertzian} for Hertzian spheres (black dots). 
}
    \label{stability_curve}
\end{figure}

\subsection{Fluid-Solid Phase Stability Analysis}\label{stability_curve_section}

To analyze the freezing/melting phase transition of a microgel suspension, we conducted a series of simulations over ranges of cross-link fraction and volume fraction for each of the three models. In each simulation, we chose the same system parameters, initializing $N_m=108$ particles on an FCC lattice, and followed the same protocol as outlined in Sec.~\ref{lindemann_ratio_section}, except that we equilibrated for $5\times 10^4$ MC steps and collected statistics (i.e., sampled configurations) over an additional $10^5$ steps to compute the equilibrium swelling ratio and Lindemann ratio. 

Fixing $x$ and starting from a concentration in the fluid phase (where $L>0.15$), we incrementally increased the dry volume fraction -- performing an independent simulation for each new value of $\phi_0$ -- until the initial crystal did not melt according to the Lindemann criterion ($L<0.15$). In the fluid phase, $L$ can grow arbitrarily large with increasing number of MC steps, while in the solid phase, $L$ maintains a steady value, implying stability (or metastability) with respect to melting. We then varied $x$ and repeated the process, in this way constructing the boundary between fluid and solid phases. It is important to emphasize that, since in each simulation the particles start on an FCC lattice, our analysis cannot distinguish between stable and metastable solid phases. 

Figure~\ref{stability_curve} shows the resulting stability curve for the Hertzian, facet, and penetration models in the plane of swollen volume fraction vs.~softness, defined as the ratio $k_B T/\varepsilon$. 
The softness quantifies particle deformability, with low values corresponding to stiff, hard-sphere-like microgels and higher values to softer, more deformable particles that can more easily facet or interpenetrate at high concentrations. 
To set the scale, as the cross-link fraction increases from $x=3\times 10^{-5}$ to $5\times 10^{-5}$,
the softness decreases from $k_B T/\varepsilon=0.0057$ to 0.0034.
Also shown in Fig.~\ref{stability_curve}, for comparison, is the thermodynamic phase boundary determined by P\`amies \textit{et al.}\cite{Pamies2009Hertzian} for Hertzian spheres based on a coexistence analysis of free energies computed via thermodynamic integration.

In all of the models considered, we find that the fluid-solid transition volume fraction monotonically increases with increasing softness. This trend is consistent with the freedom of softer particles to adapt and respond to confinement by deswelling and/or deforming before forfeiting translational entropy in the solid phase. For Hertzian spheres, which lack internal degrees of freedom, but interact through a soft elastic pair potential, our stability curve lies very close to the thermodynamic phase boundary. This proximity of the two curves is consistent with the narrow fluid-solid coexistence region charted by P\`amies \textit{et al.}\cite{Pamies2009Hertzian}

Figure~\ref{stability_curve} also reveals that Hertzian spheres crystallize at significantly higher volume fractions than microgels in either the facet or penetration models. Moreover, the gap progressively widens with increasing softness, reflecting the entropic costs to microgels of faceting and interpenetrating. In the penetration model, the stability curve is nearly vertical, almost independent of softness, and lies at a volume fraction $\phi\simeq 0.52$, close to the melting density of a hard-sphere crystal.\cite{Pusey1986,Cheng2001PhaseDiagramHS} This phase behavior demonstrates the high entropic cost of penetration in our model, consistent with the steeply repulsive effective pair potential discussed in Sec.~\ref{effective_pair_potentials} and illustrated in Fig.~\ref{fig:effective_pair_potential}. For comparison, we also computed the stability curve for $\chi=0.2$ and found virtually no change in results, indicating little influence of solvent quality on the fluid-solid transition in our models. Our stability analysis supports the conclusion suggested from pair energy considerations (Sec.~\ref{effective_pair_potentials}) that soft microgels dispersed in good solvents may tend to respond to crowding by faceting rather than interpenetrating.

\subsection{Equilibrium Structural Properties}\label{structure_section}

To further diagnose and characterize phase stability, we computed two structural properties -- the radial distribution function and the orientationally averaged static structure factor. The radial distribution function, 
\begin{equation}
g(r) = \frac{2V}{N_m^2}\sum_{i<j=1}^{N_m}\la\delta(r - |\mathbf{r}_i - \mathbf{r}_j|)\ra,
\label{gr}
\end{equation}
is a measure of pair correlations and is proportional to the conditional probability of finding a particle at distance $r$ from another particle.\cite{HansenMcDonald2013} The static structure factor, 
\begin{equation}
S(q)=1+\frac{2}{N_m}\sum_{i<j=1}^{N_m}\la\frac{\sin(qr_{ij})}{qr_{ij}}\ra,
\label{Sq}
\end{equation}
is proportional to the intensity of scattered light, for scattered wave vector magnitude $q$, and to the Fourier transform of $g(r)$. In our simulations, we computed $S(q)$ from Eq.~(\ref{Sq}) and $g(r)$ by histogramming center-to-center separation distances between pairs of particles, averaging over configurations, and smoothing fluctuations by computing a running average over five bins.

\begin{figure}[t!]
    \centering
    \includegraphics[width=1.0\linewidth]{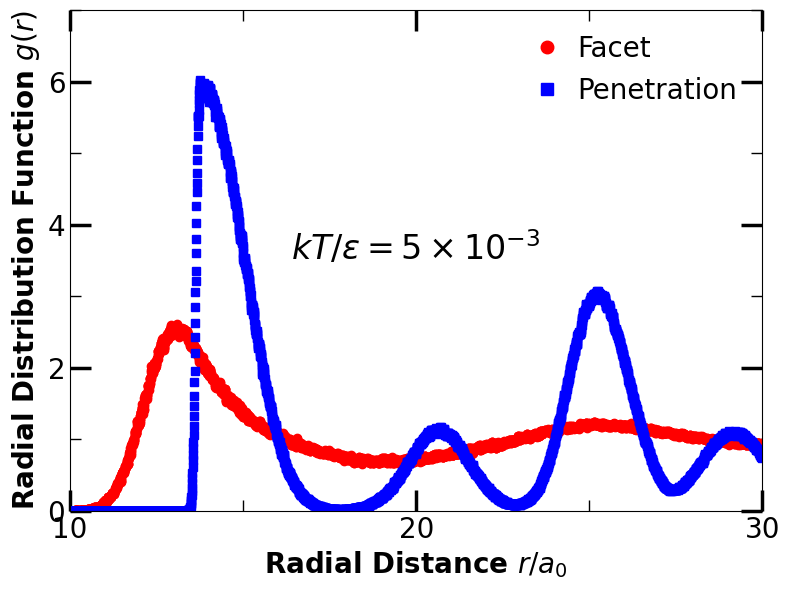}
    \caption{Simulation data for radial distribution function $g(r)$ vs.~center-to-center separation $r$ between pairs of microgels (units of dry radius) for a suspension of $N_m=108$ microgels with softness $k_BT/\varepsilon=5\times 10^{-3}$ and swollen volume fraction $\phi\simeq 0.6$.  In the facet model (red circles), the system is in a fluid state, while in the penetration model (blue squares) the solid is stable.
Other system parameters are as in Sec.~\ref{lindemann_ratio_section}. 
}
    \label{grfig}
\end{figure}
  
\begin{figure}[h!]
    \centering
    \includegraphics[width=1.0\linewidth]{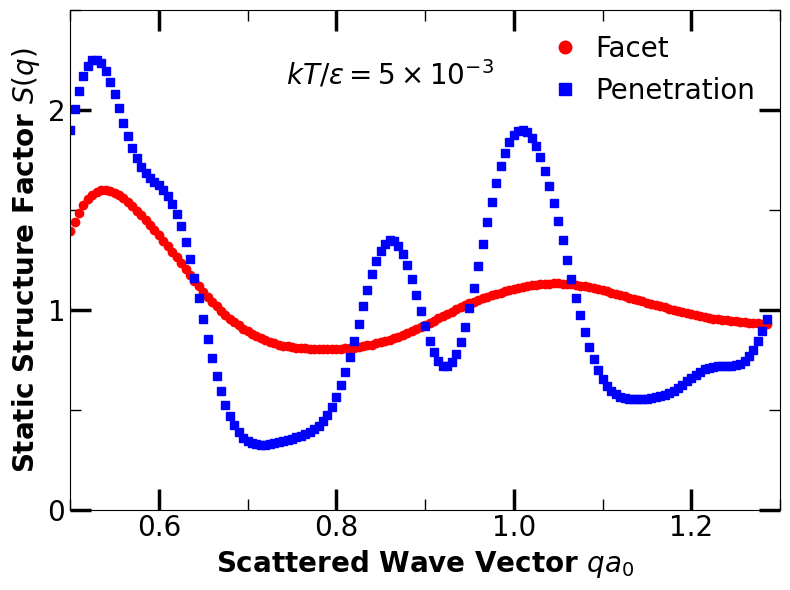}
    \caption{Simulation data for static structure factor $S(q)$ vs.~scattered wave vector $q$ (units of inverse dry radius) for a suspension of $N_m=108$ microgels with softness $k_BT/\varepsilon=5\times 10^{-3}$ and swollen volume fraction $\phi\simeq 0.6$. In the facet model (red circles), the fluid is stable, while in the penetration model (blue squares) the solid is stable. Other system parameters are as in Sec.~\ref{lindemann_ratio_section}. 
}
    \label{Sqfig}
\end{figure}

As seen in Figs.~\ref{grfig} and \ref{Sqfig}, at a point in the stability diagram where $k_BT/\varepsilon=0.005$ and $\phi\simeq 0.6$ the suspension exhibits in the facet model fluid-like structure, but in the penetration model solid-like structure, characterized by distinct peaks in both $g(r)$ and $S(q)$, commensurate with FCC neighbor distances. These structural data validate our identification of the fluid and solid phases from the Lindemann criterion in Sec.~\ref{stability_curve_section}. Note that the height of the main peak of $S(q)$ in the penetration model is somewhat lower than the freezing threshold, $S(q_{\rm max})\simeq 2.85$, predicted by the Hansen-Verlet criterion,\cite{hansen-verlet1969} which applies, however, to simple liquids and not soft particles with internal degrees of freedom.

\section{Conclusions}\label{conclusions}

In summary, we developed and analyzed coarse-grained models of compressible microgels that account for 
the freedom of these soft colloidal particles to swell/deswell, deform by faceting, and interpenetrate. 
Coupling the Flory-Rehner theory of polymer networks with the Hertz contact theory, we established 
a framework that connects cross-link fraction to both swelling behavior and interparticle interactions. 
By performing Monte Carlo simulations that incorporate trial displacements and size changes of particles,
while computing and updating particle facets and overlaps, we probed single-microgel swelling and 
deformation and their influence on the fluid-solid transition of suspensions, diagnosed by the 
Lindemann criterion, across broad ranges of microgel softness and volume fraction.

Our results for idealized microgels with homogeneous polymer and cross-link densities demonstrate that 
single-particle compressibility, deformability, and penetrability can profoundly influence phase stability 
of suspensions of soft colloidal particles. In crowded suspensions, microgels given freedom to form facets 
with their neighbors (facet model) display swelling and thermodynamic behavior strongly dependent 
on particle softness. While softer particles crystallize at relatively high volume fractions, yet still
lower than for incompressible Hertzian spheres, stiffer particles more readily crystallize. This variation 
in fluid-solid transition density highlights connections between single-particle deformation modes and 
bulk phase behavior. Validation of our facet model simulations against single-microgel confinement 
calculations confirmed the predicted swelling/deswelling trends.

The same homogeneous microgels, when allowed to respond to crowding by mutually interpenetrating 
(penetration model), exhibited steeply repulsive effective pair interactions and crystallized at much lower 
volume fractions, close to the hard-sphere fluid freezing transition, independent of cross-link fraction.
This hard-sphere-like behavior, which deviates from experimentally observed behavior, appears to support 
the prevailing view that real microgels have inhomogeneous polymer and cross-link density profiles, 
with a diffuse corona and dangling polymer chains surrounding a dense inner core. Whether our 
penetration model may better describe interactions between loosely cross-linked outer shells than 
between densely cross-linked cores can be addressed in future by adapting our models to microgels 
with core-shell architecture.

Our general prediction that internal degrees of freedom that allow microgels to compress and deform in shape 
may significantly affect thermodynamic properties of bulk suspensions compared with incompressible Hertzian 
spheres~\cite{Pamies2009Hertzian} is in qualitative agreement with the molecular-scale simulations of 
Zaccarelli {\it et al.},\cite{zaccarelli-multiHertzian2018,Rovigatti2019Hertzian,Rovigatti2019overview} 
who demonstrated that the Hertzian model fails once particle overlaps and deformation become significant. 
Also highly relevant are the molecular dynamics simulations of Del Monte and Zaccarelli,\cite{zaccarelli2024}
who showed that bead-spring microgels deswell isotropically near hard-sphere close packing, beyond which 
faceting and interpenetration become detectable. Our coarse-grained models allow microgels to respond 
to crowded environments via similar mechanisms, and we found qualitative agreement between predictions
from our facet model and their simulation data. Quantitative comparison between the two studies is 
complicated, however, by a number of factors.

First of all, Del Monte and Zaccarelli studied microgels with inhomogeneous (core-shell) 
structure, whereas our study is limited to homogeneous microgels. (We note in passing that this distinction
also limits quantitative comparisons with experiments.) Secondly, their model features an implicit 
solvent, incorporated via bead-bead effective interactions, whereas our model, although coarse-grained, 
includes polymer-solvent mixing entropy. The latter contribution is somewhat akin to the volume energy 
associated with implicit counterions included when modeling ionic microgels with effective electrostatic 
interactions.\cite{WeyerDenton2018Swelling} Finally, for simplicity, our study restricts microgels to 
respond to crowding by either faceting or interpenetrating. In future work, we plan to relax this constraint 
and combine the two mechanisms.

The findings presented here extend previous studies~\cite{Pamies2009Hertzian,MunaoSaija2019Hertzian,Fomin2018Hertzian2D,Tsiok2020Hertzian2D} of incompressible Hertzian spheres by explicitly incorporating internal degrees of freedom of microgel particles. The results underscore that softness, compressibility, and deformation pathways critically alter the effective interactions that drive crystallization in dense suspensions. Beyond providing fundamental insights, this work establishes a theoretical basis for interpreting experimental observations~\cite{PelaezFernandez2015,Hildebrandt2023FluidSolid,Conley2017Jamming,conley2019,scheffold2020pathways} of microgel suspensions and may help guide the design of responsive soft materials for applications to drug delivery, biosensing, and photonic crystals. Future work will include mapping out the thermodynamic phase diagram of compressible microgels by computing free energies from simulations via thermodynamic integration. Our approach may also be applied to microgels with different internal architectures (e.g., core-shell~\cite{zaccarelli-multiHertzian2018} or hollow~\cite{Wypysek-Scotti2019}) and may be extended to modeling the response of such soft, deformable colloids to confinement in channels of various geometries.

\section*{Acknowledgements}
This work was supported by the National Science Foundation under Grant No.~DMR-1928073.

\section*{Author Contributions}
{\bf Oreoluwa Alade:}
formal analysis (equal); methodology (equal); calculations (equal); software (equal);
data curation (equal); visualization (equal); validation (equal); writing -- original draft (equal);
writing -- review and editing (supporting).
\\[1ex]
{\bf Alan Denton:}
conceptualization (lead); project administration (lead); resources (lead); supervision (lead);
formal analysis (equal); methodology (equal); calculations (equal); software (equal);
data curation (equal); visualization (equal); validation (equal); writing -- original draft (equal);
writing -- review and editing (lead).

\section*{Conflicts of Interest}
The authors have no conflicts to disclose.

\section*{Data Availability}
All data supporting this article are included in the text and figures of the manuscript.


\balance

\bibliography{references} 

\providecommand*{\mcitethebibliography}{\thebibliography}
\csname @ifundefined\endcsname{endmcitethebibliography}
{\let\endmcitethebibliography\endthebibliography}{}
\begin{mcitethebibliography}{92}
\providecommand*{\natexlab}[1]{#1}
\providecommand*{\mciteSetBstSublistMode}[1]{}
\providecommand*{\mciteSetBstMaxWidthForm}[2]{}
\providecommand*{\mciteBstWouldAddEndPuncttrue}
  {\def\EndOfBibitem{\unskip.}}
\providecommand*{\mciteBstWouldAddEndPunctfalse}
  {\let\EndOfBibitem\relax}
\providecommand*{\mciteSetBstMidEndSepPunct}[3]{}
\providecommand*{\mciteSetBstSublistLabelBeginEnd}[3]{}
\providecommand*{\EndOfBibitem}{}
\mciteSetBstSublistMode{f}
\mciteSetBstMaxWidthForm{subitem}
{(\emph{\alph{mcitesubitemcount}})}
\mciteSetBstSublistLabelBeginEnd{\mcitemaxwidthsubitemform\space}
{\relax}{\relax}

\bibitem[Pelton(2000)]{Pelton2000}
R.~Pelton, \emph{Adv. Colloid Interface Sci.}, 2000, \textbf{85}, 1--33\relax
\mciteBstWouldAddEndPuncttrue
\mciteSetBstMidEndSepPunct{\mcitedefaultmidpunct}
{\mcitedefaultendpunct}{\mcitedefaultseppunct}\relax
\EndOfBibitem
\bibitem[Fern{\'a}ndez-Nieves \emph{et~al.}(2011)Fern{\'a}ndez-Nieves, Wyss,
  Mattsson, and Weitz]{MicrogelBook2011}
\emph{Microgel Suspensions: {F}undamentals and Applications}, ed.
  A.~Fern{\'a}ndez-Nieves, H.~Wyss, J.~Mattsson and D.~A. Weitz, Wiley-VCH
  Verlag GmbH \& Co. KGaA, Weinheim, 2011\relax
\mciteBstWouldAddEndPuncttrue
\mciteSetBstMidEndSepPunct{\mcitedefaultmidpunct}
{\mcitedefaultendpunct}{\mcitedefaultseppunct}\relax
\EndOfBibitem
\bibitem[Lyon and Serpe(2012)]{HydrogelBook2012}
\emph{Hydrogel Micro and Nanoparticles}, ed. L.~A. Lyon and M.~J. Serpe,
  Wiley-VCH Verlag GmbH \& Co. KGaA, Weinheim, 2012\relax
\mciteBstWouldAddEndPuncttrue
\mciteSetBstMidEndSepPunct{\mcitedefaultmidpunct}
{\mcitedefaultendpunct}{\mcitedefaultseppunct}\relax
\EndOfBibitem
\bibitem[Lyon and Fern{\'a}ndez-Nieves(2012)]{LyonFernandezNieves2012}
L.~A. Lyon and A.~Fern{\'a}ndez-Nieves, \emph{Annu. Rev. Phys. Chem.}, 2012,
  \textbf{63}, 25--43\relax
\mciteBstWouldAddEndPuncttrue
\mciteSetBstMidEndSepPunct{\mcitedefaultmidpunct}
{\mcitedefaultendpunct}{\mcitedefaultseppunct}\relax
\EndOfBibitem
\bibitem[Yunker \emph{et~al.}(2014)Yunker, Chen, Gratale, Lohr, Still, and
  Yodh]{Yunker2014}
P.~J. Yunker, K.~Chen, M.~D. Gratale, M.~A. Lohr, T.~Still and A.~G. Yodh,
  \emph{Rep. Prog. Phys.}, 2014, \textbf{77}, 056601\relax
\mciteBstWouldAddEndPuncttrue
\mciteSetBstMidEndSepPunct{\mcitedefaultmidpunct}
{\mcitedefaultendpunct}{\mcitedefaultseppunct}\relax
\EndOfBibitem
\bibitem[Senff and Richtering(1999)]{Senff1999}
H.~Senff and W.~Richtering, \emph{J. Chem. Phys.}, 1999, \textbf{111},
  1705--1711\relax
\mciteBstWouldAddEndPuncttrue
\mciteSetBstMidEndSepPunct{\mcitedefaultmidpunct}
{\mcitedefaultendpunct}{\mcitedefaultseppunct}\relax
\EndOfBibitem
\bibitem[Stieger \emph{et~al.}(2004)Stieger, Richtering, Pedersen, and
  Lindner]{Stieger2004}
M.~Stieger, W.~Richtering, J.~S. Pedersen and P.~Lindner, \emph{J. Chem.
  Phys.}, 2004, \textbf{120}, 6197--6206\relax
\mciteBstWouldAddEndPuncttrue
\mciteSetBstMidEndSepPunct{\mcitedefaultmidpunct}
{\mcitedefaultendpunct}{\mcitedefaultseppunct}\relax
\EndOfBibitem
\bibitem[Borrega \emph{et~al.}(1999)Borrega, Cloitre, Betremieux, Ernst, and
  Leibler]{Borrega1999}
R.~Borrega, M.~Cloitre, I.~Betremieux, B.~Ernst and L.~Leibler, \emph{Europhys.
  Lett.}, 1999, \textbf{47}, 729--735\relax
\mciteBstWouldAddEndPuncttrue
\mciteSetBstMidEndSepPunct{\mcitedefaultmidpunct}
{\mcitedefaultendpunct}{\mcitedefaultseppunct}\relax
\EndOfBibitem
\bibitem[Cloitre \emph{et~al.}(2003)Cloitre, Borrega, Monti, and
  Leibler]{Cloitre2003}
M.~Cloitre, R.~Borrega, F.~Monti and L.~Leibler, \emph{C. R. Phys.}, 2003,
  \textbf{4}, 221--230\relax
\mciteBstWouldAddEndPuncttrue
\mciteSetBstMidEndSepPunct{\mcitedefaultmidpunct}
{\mcitedefaultendpunct}{\mcitedefaultseppunct}\relax
\EndOfBibitem
\bibitem[Tan \emph{et~al.}(2004)Tan, Tam, Lam, and Tan]{Tan2004}
B.~H. Tan, K.~C. Tam, Y.~C. Lam and C.~B. Tan, \emph{J. Rheol.}, 2004,
  \textbf{48}, 915--926\relax
\mciteBstWouldAddEndPuncttrue
\mciteSetBstMidEndSepPunct{\mcitedefaultmidpunct}
{\mcitedefaultendpunct}{\mcitedefaultseppunct}\relax
\EndOfBibitem
\bibitem[Fern{\'a}ndez-Nieves \emph{et~al.}(2000)Fern{\'a}ndez-Nieves,
  Fern{\'a}ndez-Barbero, Vincent, and de~las Nieves]{FernandezNieves2000}
A.~Fern{\'a}ndez-Nieves, A.~Fern{\'a}ndez-Barbero, B.~Vincent and F.~J. de~las
  Nieves, \emph{Macromol.}, 2000, \textbf{33}, 2114--2118\relax
\mciteBstWouldAddEndPuncttrue
\mciteSetBstMidEndSepPunct{\mcitedefaultmidpunct}
{\mcitedefaultendpunct}{\mcitedefaultseppunct}\relax
\EndOfBibitem
\bibitem[Fern{\'a}ndez-Nieves \emph{et~al.}(2003)Fern{\'a}ndez-Nieves,
  Fern{\'a}ndez-Barbero, Vincent, and de~las Nieves]{FernandezNieves2003}
A.~Fern{\'a}ndez-Nieves, A.~Fern{\'a}ndez-Barbero, B.~Vincent and F.~J. de~las
  Nieves, \emph{J. Chem. Phys.}, 2003, \textbf{119}, 10383--10388\relax
\mciteBstWouldAddEndPuncttrue
\mciteSetBstMidEndSepPunct{\mcitedefaultmidpunct}
{\mcitedefaultendpunct}{\mcitedefaultseppunct}\relax
\EndOfBibitem
\bibitem[Li{\'e}tor-Santos \emph{et~al.}(2009)Li{\'e}tor-Santos,
  Sierra-Mart{\'i}n, Vavrin, Hu, Gasser, and
  Fern{\'a}ndez-Nieves]{LietorSantos2009}
J.~J. Li{\'e}tor-Santos, B.~Sierra-Mart{\'i}n, R.~Vavrin, Z.~Hu, U.~Gasser and
  A.~Fern{\'a}ndez-Nieves, \emph{Macromol.}, 2009, \textbf{42},
  6225--6230\relax
\mciteBstWouldAddEndPuncttrue
\mciteSetBstMidEndSepPunct{\mcitedefaultmidpunct}
{\mcitedefaultendpunct}{\mcitedefaultseppunct}\relax
\EndOfBibitem
\bibitem[Hertle \emph{et~al.}(2010)Hertle, Zeiser, Hasen{\"o}hrl, Busch, and
  Hellweg]{Hertle2010}
Y.~Hertle, M.~Zeiser, C.~Hasen{\"o}hrl, P.~Busch and T.~Hellweg, \emph{Colloid
  Polym. Sci.}, 2010, \textbf{288}, 1047--1059\relax
\mciteBstWouldAddEndPuncttrue
\mciteSetBstMidEndSepPunct{\mcitedefaultmidpunct}
{\mcitedefaultendpunct}{\mcitedefaultseppunct}\relax
\EndOfBibitem
\bibitem[Menut \emph{et~al.}(2012)Menut, Seiffert, Sprakel, and
  Weitz]{Menut2012}
P.~Menut, S.~Seiffert, J.~Sprakel and D.~A. Weitz, \emph{Soft Matter}, 2012,
  \textbf{8}, 156--164\relax
\mciteBstWouldAddEndPuncttrue
\mciteSetBstMidEndSepPunct{\mcitedefaultmidpunct}
{\mcitedefaultendpunct}{\mcitedefaultseppunct}\relax
\EndOfBibitem
\bibitem[Romeo \emph{et~al.}(2012)Romeo, Imperiali, Kim, Fern{\'a}ndez-Nieves,
  and Weitz]{Romeo2012}
G.~Romeo, L.~Imperiali, J.-W. Kim, A.~Fern{\'a}ndez-Nieves and D.~A. Weitz,
  \emph{J. Chem. Phys.}, 2012, \textbf{136}, 124905\relax
\mciteBstWouldAddEndPuncttrue
\mciteSetBstMidEndSepPunct{\mcitedefaultmidpunct}
{\mcitedefaultendpunct}{\mcitedefaultseppunct}\relax
\EndOfBibitem
\bibitem[Romeo and Ciamarra(2013)]{Romeo2013}
G.~Romeo and M.~P. Ciamarra, \emph{Soft Matter}, 2013, \textbf{9},
  5401--5406\relax
\mciteBstWouldAddEndPuncttrue
\mciteSetBstMidEndSepPunct{\mcitedefaultmidpunct}
{\mcitedefaultendpunct}{\mcitedefaultseppunct}\relax
\EndOfBibitem
\bibitem[Li{\'e}tor-Santos \emph{et~al.}(2011)Li{\'e}tor-Santos,
  Sierra-Mart{\'i}n, Gasser, and Fern{\'a}ndez-Nieves]{LietorSantos2011SM}
J.~J. Li{\'e}tor-Santos, B.~Sierra-Mart{\'i}n, U.~Gasser and
  A.~Fern{\'a}ndez-Nieves, \emph{Soft Matter}, 2011, \textbf{7},
  6370--6374\relax
\mciteBstWouldAddEndPuncttrue
\mciteSetBstMidEndSepPunct{\mcitedefaultmidpunct}
{\mcitedefaultendpunct}{\mcitedefaultseppunct}\relax
\EndOfBibitem
\bibitem[Sierra-Mart{\'i}n and Fern{\'a}ndez-Nieves(2012)]{SierraMartin2012}
B.~Sierra-Mart{\'i}n and A.~Fern{\'a}ndez-Nieves, \emph{Soft Matter}, 2012,
  \textbf{8}, 4141--4150\relax
\mciteBstWouldAddEndPuncttrue
\mciteSetBstMidEndSepPunct{\mcitedefaultmidpunct}
{\mcitedefaultendpunct}{\mcitedefaultseppunct}\relax
\EndOfBibitem
\bibitem[Li{\'e}tor-Santos \emph{et~al.}(2011)Li{\'e}tor-Santos,
  Sierra-Mart{\'i}n, and Fern{\'a}ndez-Nieves]{LietorSantos2011PRE}
J.~J. Li{\'e}tor-Santos, B.~Sierra-Mart{\'i}n and A.~Fern{\'a}ndez-Nieves,
  \emph{Phys. Rev. E}, 2011, \textbf{84}, 060402\relax
\mciteBstWouldAddEndPuncttrue
\mciteSetBstMidEndSepPunct{\mcitedefaultmidpunct}
{\mcitedefaultendpunct}{\mcitedefaultseppunct}\relax
\EndOfBibitem
\bibitem[Sierra-Mart{\'i}n \emph{et~al.}(2011)Sierra-Mart{\'i}n, Laporte,
  South, Lyon, and Fern{\'a}ndez-Nieves]{SierraMartin2011}
B.~Sierra-Mart{\'i}n, Y.~Laporte, A.~B. South, L.~A. Lyon and
  A.~Fern{\'a}ndez-Nieves, \emph{Phys. Rev. E}, 2011, \textbf{84}, 011406\relax
\mciteBstWouldAddEndPuncttrue
\mciteSetBstMidEndSepPunct{\mcitedefaultmidpunct}
{\mcitedefaultendpunct}{\mcitedefaultseppunct}\relax
\EndOfBibitem
\bibitem[Hashmi and Dufresne(2009)]{Hashmi2009}
S.~M. Hashmi and E.~R. Dufresne, \emph{Soft Matter}, 2009, \textbf{5},
  3682--3688\relax
\mciteBstWouldAddEndPuncttrue
\mciteSetBstMidEndSepPunct{\mcitedefaultmidpunct}
{\mcitedefaultendpunct}{\mcitedefaultseppunct}\relax
\EndOfBibitem
\bibitem[Pelaez-Fernandez \emph{et~al.}(2015)Pelaez-Fernandez, Souslov, Lyon,
  Goldbart, and Fern{\'a}ndez-Nieves]{PelaezFernandez2015}
M.~Pelaez-Fernandez, A.~Souslov, L.~A. Lyon, P.~M. Goldbart and
  A.~Fern{\'a}ndez-Nieves, \emph{Phys. Rev. Lett.}, 2015, \textbf{114},
  098303\relax
\mciteBstWouldAddEndPuncttrue
\mciteSetBstMidEndSepPunct{\mcitedefaultmidpunct}
{\mcitedefaultendpunct}{\mcitedefaultseppunct}\relax
\EndOfBibitem
\bibitem[Mason \emph{et~al.}(1995)Mason, Bibette, and Weitz]{Mason1995}
T.~G. Mason, J.~Bibette and D.~A. Weitz, \emph{Phys. Rev. Lett.}, 1995,
  \textbf{75}, 2051--2054\relax
\mciteBstWouldAddEndPuncttrue
\mciteSetBstMidEndSepPunct{\mcitedefaultmidpunct}
{\mcitedefaultendpunct}{\mcitedefaultseppunct}\relax
\EndOfBibitem
\bibitem[Gr{\"o}hn and Antonietti(2000)]{Grohn2000}
F.~Gr{\"o}hn and M.~Antonietti, \emph{Macromol.}, 2000, \textbf{33},
  5938--5949\relax
\mciteBstWouldAddEndPuncttrue
\mciteSetBstMidEndSepPunct{\mcitedefaultmidpunct}
{\mcitedefaultendpunct}{\mcitedefaultseppunct}\relax
\EndOfBibitem
\bibitem[Levin(2002)]{Levin2002}
Y.~Levin, \emph{Phys. Rev. E}, 2002, \textbf{65}, 036143\relax
\mciteBstWouldAddEndPuncttrue
\mciteSetBstMidEndSepPunct{\mcitedefaultmidpunct}
{\mcitedefaultendpunct}{\mcitedefaultseppunct}\relax
\EndOfBibitem
\bibitem[Fern{\'a}ndez-Nieves and M{\'a}rquez(2005)]{FernandezNieves2005}
A.~Fern{\'a}ndez-Nieves and M.~M{\'a}rquez, \emph{J. Chem. Phys.}, 2005,
  \textbf{122}, 084702\relax
\mciteBstWouldAddEndPuncttrue
\mciteSetBstMidEndSepPunct{\mcitedefaultmidpunct}
{\mcitedefaultendpunct}{\mcitedefaultseppunct}\relax
\EndOfBibitem
\bibitem[Singh \emph{et~al.}(2012)Singh, Fedosov, Chatterji, Winkler, and
  Gompper]{Singh2012}
S.~P. Singh, D.~A. Fedosov, A.~Chatterji, R.~G. Winkler and G.~Gompper,
  \emph{J. Phys.: Condens. Matter}, 2012, \textbf{24}, 464103\relax
\mciteBstWouldAddEndPuncttrue
\mciteSetBstMidEndSepPunct{\mcitedefaultmidpunct}
{\mcitedefaultendpunct}{\mcitedefaultseppunct}\relax
\EndOfBibitem
\bibitem[Winkler \emph{et~al.}(2014)Winkler, Fedosov, and Gompper]{Winkler2014}
R.~G. Winkler, D.~A. Fedosov and G.~Gompper, \emph{Curr. Opin. Colloid
  Interface Sci.}, 2014, \textbf{19}, 594--610\relax
\mciteBstWouldAddEndPuncttrue
\mciteSetBstMidEndSepPunct{\mcitedefaultmidpunct}
{\mcitedefaultendpunct}{\mcitedefaultseppunct}\relax
\EndOfBibitem
\bibitem[Li \emph{et~al.}(2014)Li, S{\'a}nchez-D{\'i}az, Wu, Hamilton, Falus,
  Porcar, Liu, Do, Faraone, Smith, Egami, and Chen]{Li2014}
X.~Li, L.~E. S{\'a}nchez-D{\'i}az, B.~Wu, W.~A. Hamilton, P.~Falus, L.~Porcar,
  Y.~Liu, C.~Do, A.~Faraone, G.~S. Smith, T.~Egami and W.-R. Chen, \emph{ACS
  Macro Lett.}, 2014, \textbf{3}, 1271--1275\relax
\mciteBstWouldAddEndPuncttrue
\mciteSetBstMidEndSepPunct{\mcitedefaultmidpunct}
{\mcitedefaultendpunct}{\mcitedefaultseppunct}\relax
\EndOfBibitem
\bibitem[Egorov \emph{et~al.}(2013)Egorov, Paturej, Likos, and
  Milchev]{Egorov2013}
S.~A. Egorov, J.~Paturej, C.~N. Likos and A.~Milchev, \emph{Macromol.}, 2013,
  \textbf{46}, 3648--3653\relax
\mciteBstWouldAddEndPuncttrue
\mciteSetBstMidEndSepPunct{\mcitedefaultmidpunct}
{\mcitedefaultendpunct}{\mcitedefaultseppunct}\relax
\EndOfBibitem
\bibitem[Colla \emph{et~al.}(2014)Colla, Likos, and Levin]{Colla2014}
T.~Colla, C.~N. Likos and Y.~Levin, \emph{J. Chem. Phys.}, 2014, \textbf{141},
  234902\relax
\mciteBstWouldAddEndPuncttrue
\mciteSetBstMidEndSepPunct{\mcitedefaultmidpunct}
{\mcitedefaultendpunct}{\mcitedefaultseppunct}\relax
\EndOfBibitem
\bibitem[Colla and Likos(2015)]{Colla2015}
T.~Colla and C.~N. Likos, \emph{Mol. Phys.}, 2015, \textbf{113},
  2496--2510\relax
\mciteBstWouldAddEndPuncttrue
\mciteSetBstMidEndSepPunct{\mcitedefaultmidpunct}
{\mcitedefaultendpunct}{\mcitedefaultseppunct}\relax
\EndOfBibitem
\bibitem[Gupta \emph{et~al.}(2015)Gupta, Camargo, Stellbrink, Allgaier,
  Radulescu, Lindner, Zaccarelli, Likos, and Richter]{Gupta2015Nanoscale}
S.~Gupta, M.~Camargo, J.~Stellbrink, J.~Allgaier, A.~Radulescu, P.~Lindner,
  E.~Zaccarelli, C.~N. Likos and D.~Richter, \emph{Nanoscale}, 2015,
  \textbf{7}, 13924--13934\relax
\mciteBstWouldAddEndPuncttrue
\mciteSetBstMidEndSepPunct{\mcitedefaultmidpunct}
{\mcitedefaultendpunct}{\mcitedefaultseppunct}\relax
\EndOfBibitem
\bibitem[Gupta \emph{et~al.}(2015)Gupta, Stellbrink, Zaccarelli, Likos,
  Camargo, Holmqvist, Allgaier, Willner, and Richter]{Gupta2015PRL}
S.~Gupta, J.~Stellbrink, E.~Zaccarelli, C.~N. Likos, M.~Camargo, P.~Holmqvist,
  J.~Allgaier, L.~Willner and D.~Richter, \emph{Phys. Rev. Lett.}, 2015,
  \textbf{115}, 128302\relax
\mciteBstWouldAddEndPuncttrue
\mciteSetBstMidEndSepPunct{\mcitedefaultmidpunct}
{\mcitedefaultendpunct}{\mcitedefaultseppunct}\relax
\EndOfBibitem
\bibitem[Scotti \emph{et~al.}(2022)Scotti, Schulte, Lopez, Crassous, Bochenek,
  and Richtering]{scotti2022}
A.~Scotti, M.~F. Schulte, C.~G. Lopez, J.~J. Crassous, S.~Bochenek and
  W.~Richtering, \emph{Chem. Rev.}, 2022, \textbf{122}, 11675--11700\relax
\mciteBstWouldAddEndPuncttrue
\mciteSetBstMidEndSepPunct{\mcitedefaultmidpunct}
{\mcitedefaultendpunct}{\mcitedefaultseppunct}\relax
\EndOfBibitem
\bibitem[P{\'a}mies \emph{et~al.}(2009)P{\'a}mies, Cacciuto, and
  Frenkel]{Pamies2009Hertzian}
J.~C. P{\'a}mies, A.~Cacciuto and D.~Frenkel, \emph{J. Chem. Phys.}, 2009,
  \textbf{131}, 044514\relax
\mciteBstWouldAddEndPuncttrue
\mciteSetBstMidEndSepPunct{\mcitedefaultmidpunct}
{\mcitedefaultendpunct}{\mcitedefaultseppunct}\relax
\EndOfBibitem
\bibitem[Likos \emph{et~al.}(1998)Likos, Watzlawek, and
  L{\"o}wen]{Likos1998Penetrable}
C.~N. Likos, M.~Watzlawek and H.~L{\"o}wen, \emph{Phys. Rev. E}, 1998,
  \textbf{58}, 3135--3144\relax
\mciteBstWouldAddEndPuncttrue
\mciteSetBstMidEndSepPunct{\mcitedefaultmidpunct}
{\mcitedefaultendpunct}{\mcitedefaultseppunct}\relax
\EndOfBibitem
\bibitem[Hildebrandt \emph{et~al.}(2023)Hildebrandt, Thuy, Kippenberger,
  Wigger, Houston, Scotti, and Karg]{Hildebrandt2023FluidSolid}
M.~Hildebrandt, D.~P. Thuy, J.~Kippenberger, T.~L. Wigger, J.~E. Houston,
  A.~Scotti and M.~Karg, \emph{Soft Matter}, 2023, \textbf{19},
  7122--7135\relax
\mciteBstWouldAddEndPuncttrue
\mciteSetBstMidEndSepPunct{\mcitedefaultmidpunct}
{\mcitedefaultendpunct}{\mcitedefaultseppunct}\relax
\EndOfBibitem
\bibitem[Conley \emph{et~al.}(2017)Conley, Aebischer, N{\"o}jd, Schurtenberger,
  and Scheffold]{Conley2017Jamming}
G.~M. Conley, P.~Aebischer, S.~N{\"o}jd, P.~Schurtenberger and F.~Scheffold,
  \emph{Sci. Adv.}, 2017, \textbf{3}, e1700969\relax
\mciteBstWouldAddEndPuncttrue
\mciteSetBstMidEndSepPunct{\mcitedefaultmidpunct}
{\mcitedefaultendpunct}{\mcitedefaultseppunct}\relax
\EndOfBibitem
\bibitem[Conley \emph{et~al.}(2019)Conley, Zhang, Aebischer, Harden, and
  Scheffold]{conley2019}
G.~M. Conley, C.~Zhang, P.~Aebischer, J.~L. Harden and F.~Scheffold, \emph{Nat.
  Commun.}, 2019, \textbf{10}, 1--8\relax
\mciteBstWouldAddEndPuncttrue
\mciteSetBstMidEndSepPunct{\mcitedefaultmidpunct}
{\mcitedefaultendpunct}{\mcitedefaultseppunct}\relax
\EndOfBibitem
\bibitem[Scheffold(2020)]{scheffold2020pathways}
F.~Scheffold, \emph{Nat. Commun.}, 2020, \textbf{11}, 4315\relax
\mciteBstWouldAddEndPuncttrue
\mciteSetBstMidEndSepPunct{\mcitedefaultmidpunct}
{\mcitedefaultendpunct}{\mcitedefaultseppunct}\relax
\EndOfBibitem
\bibitem[Meijer \emph{et~al.}(2018)Meijer, Rovigatti, Zaccarelli, and
  Schurtenberger]{zaccarelli-multiHertzian2018}
J.-M. Meijer, L.~Rovigatti, E.~Zaccarelli and P.~Schurtenberger, \emph{Nat.
  Commun.}, 2018, \textbf{9}, 5039\relax
\mciteBstWouldAddEndPuncttrue
\mciteSetBstMidEndSepPunct{\mcitedefaultmidpunct}
{\mcitedefaultendpunct}{\mcitedefaultseppunct}\relax
\EndOfBibitem
\bibitem[Rovigatti \emph{et~al.}(2019)Rovigatti, Gnan, Ninarello, and
  Zaccarelli]{Rovigatti2019Hertzian}
L.~Rovigatti, N.~Gnan, A.~Ninarello and E.~Zaccarelli, \emph{Macromol.}, 2019,
  \textbf{52}, 4895--4906\relax
\mciteBstWouldAddEndPuncttrue
\mciteSetBstMidEndSepPunct{\mcitedefaultmidpunct}
{\mcitedefaultendpunct}{\mcitedefaultseppunct}\relax
\EndOfBibitem
\bibitem[Rovigatti \emph{et~al.}(2019)Rovigatti, Gnan, Tavagnacco, Moreno, and
  Zaccarelli]{Rovigatti2019overview}
L.~Rovigatti, N.~Gnan, L.~Tavagnacco, A.~J. Moreno and E.~Zaccarelli,
  \emph{Soft Matter}, 2019, \textbf{15}, 1108--1119\relax
\mciteBstWouldAddEndPuncttrue
\mciteSetBstMidEndSepPunct{\mcitedefaultmidpunct}
{\mcitedefaultendpunct}{\mcitedefaultseppunct}\relax
\EndOfBibitem
\bibitem[Del~Monte and Zaccarelli(2024)]{zaccarelli2024}
G.~Del~Monte and E.~Zaccarelli, \emph{Phys. Rev. X}, 2024, \textbf{14},
  041067\relax
\mciteBstWouldAddEndPuncttrue
\mciteSetBstMidEndSepPunct{\mcitedefaultmidpunct}
{\mcitedefaultendpunct}{\mcitedefaultseppunct}\relax
\EndOfBibitem
\bibitem[Frenkel and Smit(2023)]{FrenkelSmit2023}
D.~Frenkel and B.~Smit, \emph{Understanding Molecular Simulation: {F}rom
  Algorithms to Applications}, Elsevier, 3rd edn., 2023\relax
\mciteBstWouldAddEndPuncttrue
\mciteSetBstMidEndSepPunct{\mcitedefaultmidpunct}
{\mcitedefaultendpunct}{\mcitedefaultseppunct}\relax
\EndOfBibitem
\bibitem[Frenkel and Ladd(1984)]{FrenkelLadd1984}
D.~Frenkel and A.~J.~C. Ladd, \emph{J. Chem. Phys.}, 1984, \textbf{81},
  3188--3193\relax
\mciteBstWouldAddEndPuncttrue
\mciteSetBstMidEndSepPunct{\mcitedefaultmidpunct}
{\mcitedefaultendpunct}{\mcitedefaultseppunct}\relax
\EndOfBibitem
\bibitem[Muna{\`o} and Saija(2019)]{MunaoSaija2019Hertzian}
G.~Muna{\`o} and F.~Saija, \emph{J. Chem. Phys.}, 2019, \textbf{151},
  134901\relax
\mciteBstWouldAddEndPuncttrue
\mciteSetBstMidEndSepPunct{\mcitedefaultmidpunct}
{\mcitedefaultendpunct}{\mcitedefaultseppunct}\relax
\EndOfBibitem
\bibitem[Fomin \emph{et~al.}(2018)Fomin, Gaiduk, Tsiok, and
  Ryzhov]{Fomin2018Hertzian2D}
Y.~D. Fomin, E.~A. Gaiduk, E.~N. Tsiok and V.~N. Ryzhov, \emph{Mol. Phys.},
  2018, \textbf{116}, 3258--3270\relax
\mciteBstWouldAddEndPuncttrue
\mciteSetBstMidEndSepPunct{\mcitedefaultmidpunct}
{\mcitedefaultendpunct}{\mcitedefaultseppunct}\relax
\EndOfBibitem
\bibitem[Tsiok \emph{et~al.}(2020)Tsiok, Gaiduk, Fomin, and
  Ryzhov]{Tsiok2020Hertzian2D}
E.~N. Tsiok, E.~A. Gaiduk, Y.~D. Fomin and V.~N. Ryzhov, \emph{Soft Matter},
  2020, \textbf{16}, 3962--3972\relax
\mciteBstWouldAddEndPuncttrue
\mciteSetBstMidEndSepPunct{\mcitedefaultmidpunct}
{\mcitedefaultendpunct}{\mcitedefaultseppunct}\relax
\EndOfBibitem
\bibitem[Moncho-Jord{\'a} \emph{et~al.}(2025)Moncho-Jord{\'a}, Cuetos,
  Fernandez-Rodriguez, Dzubiella, and Patti]{moncho-jorda2025}
A.~Moncho-Jord{\'a}, A.~Cuetos, M.~A. Fernandez-Rodriguez, J.~Dzubiella and
  A.~Patti, \emph{Macromol.}, 2025, \textbf{58}, 10659--10676\relax
\mciteBstWouldAddEndPuncttrue
\mciteSetBstMidEndSepPunct{\mcitedefaultmidpunct}
{\mcitedefaultendpunct}{\mcitedefaultseppunct}\relax
\EndOfBibitem
\bibitem[Pelton and Hoare(2011)]{pelton2011}
R.~Pelton and T.~Hoare, in \emph{Microgel Suspensions: {F}undamentals and
  Applications}, ed. A.~Fern{\'a}ndez-Nieves, H.~Wyss, J.~Mattsson and D.~A.
  Weitz, Wiley-VCH Verlag GmbH \& Co. KGaA, Weinheim, 2011, pp. 3--32\relax
\mciteBstWouldAddEndPuncttrue
\mciteSetBstMidEndSepPunct{\mcitedefaultmidpunct}
{\mcitedefaultendpunct}{\mcitedefaultseppunct}\relax
\EndOfBibitem
\bibitem[Acciaro \emph{et~al.}(2011)Acciaro, Gil{\'a}nyi, and Varga]{varga2011}
R.~Acciaro, T.~Gil{\'a}nyi and I.~Varga, \emph{Langmuir}, 2011, \textbf{27},
  7917--7925\relax
\mciteBstWouldAddEndPuncttrue
\mciteSetBstMidEndSepPunct{\mcitedefaultmidpunct}
{\mcitedefaultendpunct}{\mcitedefaultseppunct}\relax
\EndOfBibitem
\bibitem[Karg \emph{et~al.}(2019)Karg, Pich, Hellweg, Hoare, Lyon, Crassous,
  Suzuki, Gumerov, Schneider, Potemkin, and Richtering]{karg-richtering2019}
M.~Karg, A.~Pich, T.~Hellweg, T.~Hoare, L.~A. Lyon, J.~J. Crassous, D.~Suzuki,
  R.~A. Gumerov, S.~Schneider, I.~I. Potemkin and W.~Richtering,
  \emph{Langmuir}, 2019, \textbf{35}, 6231--6255\relax
\mciteBstWouldAddEndPuncttrue
\mciteSetBstMidEndSepPunct{\mcitedefaultmidpunct}
{\mcitedefaultendpunct}{\mcitedefaultseppunct}\relax
\EndOfBibitem
\bibitem[Urich and Denton(2016)]{UrichDenton2016}
M.~Urich and A.~R. Denton, \emph{Soft Matter}, 2016, \textbf{12},
  9086--9094\relax
\mciteBstWouldAddEndPuncttrue
\mciteSetBstMidEndSepPunct{\mcitedefaultmidpunct}
{\mcitedefaultendpunct}{\mcitedefaultseppunct}\relax
\EndOfBibitem
\bibitem[Weyer and Denton(2018)]{WeyerDenton2018Swelling}
T.~J. Weyer and A.~R. Denton, \emph{Soft Matter}, 2018, \textbf{14},
  4530--4540\relax
\mciteBstWouldAddEndPuncttrue
\mciteSetBstMidEndSepPunct{\mcitedefaultmidpunct}
{\mcitedefaultendpunct}{\mcitedefaultseppunct}\relax
\EndOfBibitem
\bibitem[Flory and Rehner(1943)]{FloryRehner1943I}
P.~J. Flory and J.~Rehner, \emph{J. Chem. Phys.}, 1943, \textbf{11},
  512--520\relax
\mciteBstWouldAddEndPuncttrue
\mciteSetBstMidEndSepPunct{\mcitedefaultmidpunct}
{\mcitedefaultendpunct}{\mcitedefaultseppunct}\relax
\EndOfBibitem
\bibitem[Flory and Rehner(1943)]{FloryRehner1943II}
P.~J. Flory and J.~Rehner, \emph{J. Chem. Phys.}, 1943, \textbf{11},
  521--526\relax
\mciteBstWouldAddEndPuncttrue
\mciteSetBstMidEndSepPunct{\mcitedefaultmidpunct}
{\mcitedefaultendpunct}{\mcitedefaultseppunct}\relax
\EndOfBibitem
\bibitem[Flory(1953)]{Flory1953Principles}
P.~J. Flory, \emph{Principles of Polymer Chemistry}, Cornell University Press,
  Ithaca, 1953\relax
\mciteBstWouldAddEndPuncttrue
\mciteSetBstMidEndSepPunct{\mcitedefaultmidpunct}
{\mcitedefaultendpunct}{\mcitedefaultseppunct}\relax
\EndOfBibitem
\bibitem[Riest \emph{et~al.}(2015)Riest, Athanasopoulou, Egorov, Likos, and
  Ziherl]{Riest2015SciRep}
J.~Riest, L.~Athanasopoulou, S.~A. Egorov, C.~N. Likos and P.~Ziherl,
  \emph{Sci. Rep.}, 2015, \textbf{5}, 15854\relax
\mciteBstWouldAddEndPuncttrue
\mciteSetBstMidEndSepPunct{\mcitedefaultmidpunct}
{\mcitedefaultendpunct}{\mcitedefaultseppunct}\relax
\EndOfBibitem
\bibitem[Doi(2013)]{doi2013soft}
M.~Doi, \emph{Soft Matter Physics}, Oxford University Press, 2013\relax
\mciteBstWouldAddEndPuncttrue
\mciteSetBstMidEndSepPunct{\mcitedefaultmidpunct}
{\mcitedefaultendpunct}{\mcitedefaultseppunct}\relax
\EndOfBibitem
\bibitem[Rubinstein and Colby(2023)]{rubinstein-colby2023}
M.~Rubinstein and R.~Colby, \emph{Polymer Physics}, Oxford University Press,
  2023\relax
\mciteBstWouldAddEndPuncttrue
\mciteSetBstMidEndSepPunct{\mcitedefaultmidpunct}
{\mcitedefaultendpunct}{\mcitedefaultseppunct}\relax
\EndOfBibitem
\bibitem[Rumyantsev \emph{et~al.}(2015)Rumyantsev, Rudov, and
  Potemkin]{Rumyantsev2015}
A.~M. Rumyantsev, A.~A. Rudov and I.~I. Potemkin, \emph{J. Chem. Phys.}, 2015,
  \textbf{142}, 171105\relax
\mciteBstWouldAddEndPuncttrue
\mciteSetBstMidEndSepPunct{\mcitedefaultmidpunct}
{\mcitedefaultendpunct}{\mcitedefaultseppunct}\relax
\EndOfBibitem
\bibitem[Saunders \emph{et~al.}(2009)Saunders, Laajam, Daly, Teow, Hu, and
  Stepto]{Saunders2009}
B.~R. Saunders, N.~Laajam, E.~Daly, S.~Teow, X.~Hu and R.~Stepto, \emph{Adv.
  Colloid Interface Sci.}, 2009, \textbf{147--148}, 251--262\relax
\mciteBstWouldAddEndPuncttrue
\mciteSetBstMidEndSepPunct{\mcitedefaultmidpunct}
{\mcitedefaultendpunct}{\mcitedefaultseppunct}\relax
\EndOfBibitem
\bibitem[Riest \emph{et~al.}(2012)Riest, Mohanty, Schurtenberger, and
  Likos]{Riest2012}
J.~Riest, P.~Mohanty, P.~Schurtenberger and C.~N. Likos, \emph{Z. Phys. Chem.},
  2012, \textbf{226}, 711--735\relax
\mciteBstWouldAddEndPuncttrue
\mciteSetBstMidEndSepPunct{\mcitedefaultmidpunct}
{\mcitedefaultendpunct}{\mcitedefaultseppunct}\relax
\EndOfBibitem
\bibitem[Lopez and Richtering(2017)]{LopezRichtering2017}
C.~G. Lopez and W.~Richtering, \emph{Soft Matter}, 2017, \textbf{13},
  8271--8280\relax
\mciteBstWouldAddEndPuncttrue
\mciteSetBstMidEndSepPunct{\mcitedefaultmidpunct}
{\mcitedefaultendpunct}{\mcitedefaultseppunct}\relax
\EndOfBibitem
\bibitem[Blundell and Terentjev(2009)]{BlundellTerentjev2009}
J.~R. Blundell and E.~M. Terentjev, \emph{Macromol.}, 2009, \textbf{42},
  5388--5394\relax
\mciteBstWouldAddEndPuncttrue
\mciteSetBstMidEndSepPunct{\mcitedefaultmidpunct}
{\mcitedefaultendpunct}{\mcitedefaultseppunct}\relax
\EndOfBibitem
\bibitem[Roa \emph{et~al.}(2017)Roa, Kim, Kandu{\v c}, Dzubiella, and
  Angioletti-Uberti]{Roa2017Catalysis}
R.~Roa, W.~K. Kim, M.~Kandu{\v c}, J.~Dzubiella and S.~Angioletti-Uberti,
  \emph{ACS Catal.}, 2017, \textbf{7}, 5604--5611\relax
\mciteBstWouldAddEndPuncttrue
\mciteSetBstMidEndSepPunct{\mcitedefaultmidpunct}
{\mcitedefaultendpunct}{\mcitedefaultseppunct}\relax
\EndOfBibitem
\bibitem[Hertz(1881)]{Hertz1881contact}
H.~Hertz, \emph{J. reine angewandte Mathematik}, 1881, \textbf{92},
  156--171\relax
\mciteBstWouldAddEndPuncttrue
\mciteSetBstMidEndSepPunct{\mcitedefaultmidpunct}
{\mcitedefaultendpunct}{\mcitedefaultseppunct}\relax
\EndOfBibitem
\bibitem[Hertz(1896)]{Hertz1896contact}
H.~Hertz, in \emph{Miscellaneous Papers}, ed. P.~Lenard, Macmillan, London,
  1896, pp. 146--162\relax
\mciteBstWouldAddEndPuncttrue
\mciteSetBstMidEndSepPunct{\mcitedefaultmidpunct}
{\mcitedefaultendpunct}{\mcitedefaultseppunct}\relax
\EndOfBibitem
\bibitem[Hertz(1882)]{Hertz1882hardness}
H.~Hertz, \emph{Verhandlungen des Vereins zur Bef{\"o}rderung des
  Gewerbefleisses}, 1882,  449--463\relax
\mciteBstWouldAddEndPuncttrue
\mciteSetBstMidEndSepPunct{\mcitedefaultmidpunct}
{\mcitedefaultendpunct}{\mcitedefaultseppunct}\relax
\EndOfBibitem
\bibitem[Hertz(1896)]{Hertz1896hardness}
H.~Hertz, in \emph{Miscellaneous Papers}, ed. P.~Lenard, Macmillan, London,
  1896, pp. 163--183\relax
\mciteBstWouldAddEndPuncttrue
\mciteSetBstMidEndSepPunct{\mcitedefaultmidpunct}
{\mcitedefaultendpunct}{\mcitedefaultseppunct}\relax
\EndOfBibitem
\bibitem[Landau and Lifshitz(1986)]{LandauLifshitzElasticity1986}
L.~D. Landau and E.~M. Lifshitz, \emph{Theory of Elasticity}, Elsevier,
  Amsterdam, 3rd edn., 1986, vol.~7\relax
\mciteBstWouldAddEndPuncttrue
\mciteSetBstMidEndSepPunct{\mcitedefaultmidpunct}
{\mcitedefaultendpunct}{\mcitedefaultseppunct}\relax
\EndOfBibitem
\bibitem[Audoly and Pomeau(2010)]{AudolyPomeau2010}
B.~Audoly and Y.~Pomeau, \emph{Elasticity and Geometry: {F}rom Hair Curls to
  the Nonlinear Response of Shells}, Oxford, New York, 2010, pp. 492--494\relax
\mciteBstWouldAddEndPuncttrue
\mciteSetBstMidEndSepPunct{\mcitedefaultmidpunct}
{\mcitedefaultendpunct}{\mcitedefaultseppunct}\relax
\EndOfBibitem
\bibitem[Geissler and Hecht(1981)]{geissler1981}
E.~Geissler and A.~M. Hecht, \emph{Macromol.}, 1981, \textbf{14},
  185--188\relax
\mciteBstWouldAddEndPuncttrue
\mciteSetBstMidEndSepPunct{\mcitedefaultmidpunct}
{\mcitedefaultendpunct}{\mcitedefaultseppunct}\relax
\EndOfBibitem
\bibitem[Li \emph{et~al.}(1993)Li, Hu, and Li]{li1993}
C.~Li, Z.~Hu and Y.~Li, \emph{Phys. Rev. E}, 1993, \textbf{48}, 603--606\relax
\mciteBstWouldAddEndPuncttrue
\mciteSetBstMidEndSepPunct{\mcitedefaultmidpunct}
{\mcitedefaultendpunct}{\mcitedefaultseppunct}\relax
\EndOfBibitem
\bibitem[Matzelle \emph{et~al.}(2003)Matzelle, Geuskens, and
  Kruse]{matzelle2003}
T.~R. Matzelle, G.~Geuskens and N.~Kruse, \emph{Macromol.}, 2003, \textbf{36},
  2926--2931\relax
\mciteBstWouldAddEndPuncttrue
\mciteSetBstMidEndSepPunct{\mcitedefaultmidpunct}
{\mcitedefaultendpunct}{\mcitedefaultseppunct}\relax
\EndOfBibitem
\bibitem[de~Gennes(1979)]{deGennes1979Scaling}
P.~G. de~Gennes, \emph{Scaling Concepts in Polymer Physics}, Cornell University
  Press, Ithaca, 1979\relax
\mciteBstWouldAddEndPuncttrue
\mciteSetBstMidEndSepPunct{\mcitedefaultmidpunct}
{\mcitedefaultendpunct}{\mcitedefaultseppunct}\relax
\EndOfBibitem
\bibitem[Wool(2005)]{FloryHugginsMixing}
R.~P. Wool, in \emph{Bio-Based Polymers and Composites}, ed. R.~P. Wool and
  X.~S. Sun, Academic Press, Burlington, 2005, pp. 551--598\relax
\mciteBstWouldAddEndPuncttrue
\mciteSetBstMidEndSepPunct{\mcitedefaultmidpunct}
{\mcitedefaultendpunct}{\mcitedefaultseppunct}\relax
\EndOfBibitem
\bibitem[Gould \emph{et~al.}(2007)Gould, Tobochnik, and Christian]{osp-sip2007}
H.~Gould, J.~Tobochnik and W.~Christian, \emph{Introduction to Computer
  Simulation Methods}, Addison Wesley, 3rd edn., 2007\relax
\mciteBstWouldAddEndPuncttrue
\mciteSetBstMidEndSepPunct{\mcitedefaultmidpunct}
{\mcitedefaultendpunct}{\mcitedefaultseppunct}\relax
\EndOfBibitem
\bibitem[Christian(2006)]{osp2006}
W.~Christian, \emph{{Open Source Physics: {A} User's Guide with Examples}},
  Addison Wesley, 2006\relax
\mciteBstWouldAddEndPuncttrue
\mciteSetBstMidEndSepPunct{\mcitedefaultmidpunct}
{\mcitedefaultendpunct}{\mcitedefaultseppunct}\relax
\EndOfBibitem
\bibitem[Binder and Heermann(2010)]{BinderHeermann2010}
K.~Binder and D.~W. Heermann, \emph{Monte Carlo Simulation in Statistical
  Physics: {A}n Introduction}, Springer, Berlin, Heidelberg, 5th edn.,
  2010\relax
\mciteBstWouldAddEndPuncttrue
\mciteSetBstMidEndSepPunct{\mcitedefaultmidpunct}
{\mcitedefaultendpunct}{\mcitedefaultseppunct}\relax
\EndOfBibitem
\bibitem[Lindemann(1910)]{Lindemann1910}
F.~A. Lindemann, \emph{Phys. Z.}, 1910, \textbf{11}, 609--612\relax
\mciteBstWouldAddEndPuncttrue
\mciteSetBstMidEndSepPunct{\mcitedefaultmidpunct}
{\mcitedefaultendpunct}{\mcitedefaultseppunct}\relax
\EndOfBibitem
\bibitem[Hoover and Ree(1968)]{HooverRee1968}
W.~G. Hoover and F.~H. Ree, \emph{J. Chem. Phys.}, 1968, \textbf{49},
  3609--3617\relax
\mciteBstWouldAddEndPuncttrue
\mciteSetBstMidEndSepPunct{\mcitedefaultmidpunct}
{\mcitedefaultendpunct}{\mcitedefaultseppunct}\relax
\EndOfBibitem
\bibitem[Hoffmann(2004)]{Hoffmann2004OnLM}
H.~J. Hoffmann, \emph{Mater. Werkst.}, 2004, \textbf{35}, 226--229\relax
\mciteBstWouldAddEndPuncttrue
\mciteSetBstMidEndSepPunct{\mcitedefaultmidpunct}
{\mcitedefaultendpunct}{\mcitedefaultseppunct}\relax
\EndOfBibitem
\bibitem[Khrapak(2020)]{Khrapak2020}
S.~Khrapak, \emph{Phys. Rev. Res.}, 2020, \textbf{2}, 012040\relax
\mciteBstWouldAddEndPuncttrue
\mciteSetBstMidEndSepPunct{\mcitedefaultmidpunct}
{\mcitedefaultendpunct}{\mcitedefaultseppunct}\relax
\EndOfBibitem
\bibitem[Pusey and van Megen(1986)]{Pusey1986}
P.~N. Pusey and W.~van Megen, \emph{Nature}, 1986, \textbf{320}, 340--342\relax
\mciteBstWouldAddEndPuncttrue
\mciteSetBstMidEndSepPunct{\mcitedefaultmidpunct}
{\mcitedefaultendpunct}{\mcitedefaultseppunct}\relax
\EndOfBibitem
\bibitem[Cheng \emph{et~al.}(2001)Cheng, Chaikin, Russel, Meyer, Zhu, Rogers,
  and Ottewill]{Cheng2001PhaseDiagramHS}
Z.~Cheng, P.~M. Chaikin, W.~B. Russel, W.~V. Meyer, J.~Zhu, R.~B. Rogers and
  R.~H. Ottewill, \emph{Mater. Des.}, 2001, \textbf{22}, 529--534\relax
\mciteBstWouldAddEndPuncttrue
\mciteSetBstMidEndSepPunct{\mcitedefaultmidpunct}
{\mcitedefaultendpunct}{\mcitedefaultseppunct}\relax
\EndOfBibitem
\bibitem[Hansen and McDonald(2013)]{HansenMcDonald2013}
J.-P. Hansen and I.~R. McDonald, \emph{Theory of Simple Liquids}, Academic
  Press, Oxford, 4th edn., 2013\relax
\mciteBstWouldAddEndPuncttrue
\mciteSetBstMidEndSepPunct{\mcitedefaultmidpunct}
{\mcitedefaultendpunct}{\mcitedefaultseppunct}\relax
\EndOfBibitem
\bibitem[Hansen and Verlet(1969)]{hansen-verlet1969}
J.~P. Hansen and L.~Verlet, \emph{Phys. Rev.}, 1969, \textbf{184}, 151\relax
\mciteBstWouldAddEndPuncttrue
\mciteSetBstMidEndSepPunct{\mcitedefaultmidpunct}
{\mcitedefaultendpunct}{\mcitedefaultseppunct}\relax
\EndOfBibitem
\bibitem[Wypysek \emph{et~al.}(2019)Wypysek, Scotti, Alziyadi, Potemkin,
  Denton, and Richtering]{Wypysek-Scotti2019}
S.~K. Wypysek, A.~Scotti, M.~O. Alziyadi, I.~I. Potemkin, A.~R. Denton and
  W.~Richtering, \emph{Macromol. Rapid Commun.}, 2019, \textbf{41},
  1900422--1--9\relax
\mciteBstWouldAddEndPuncttrue
\mciteSetBstMidEndSepPunct{\mcitedefaultmidpunct}
{\mcitedefaultendpunct}{\mcitedefaultseppunct}\relax
\EndOfBibitem
\end{mcitethebibliography}
\bibliographystyle{rsc} 

\end{document}